\documentclass[aps,prb,reprint,twocolumn,superscriptaddress,amsmath,amssymb,amsfonts]{revtex4-2}

\AtBeginDocument{\usepackage{booktabs}}               
\makeatletter
\g@addto@macro\bfseries{\boldmath}
\makeatother

\usepackage[varg]{txfonts}

\usepackage{changes}
\definechangesauthor[name={JV}, color=blue]{JV}

\usepackage{color}
\usepackage{hyperref}
\definecolor{cL}{RGB}{32,145,140}
\hypersetup{colorlinks=true,linkcolor=cL,citecolor=cL,urlcolor=cL} 
\usepackage{multirow}
\usepackage{amsmath}
\usepackage{graphicx}
\usepackage[percent]{overpic}
\usepackage{mathrsfs}
\usepackage{bm}
\usepackage{braket}
\usepackage[nolist,nohyperlinks]{acronym}
\usepackage{natbib}
\usepackage{microtype}
\usepackage{float}
\usepackage{mathtools}
\usepackage{soul}

\renewcommand{\AA}{\r{A}}

\renewcommand{\vec}[1]{\boldsymbol{#1}}

\newcommand{\rucl}{$\alpha$-RuCl$_3$}

\begin{document}
\title{Experimental Evidence for Non-spherical Magnetic Form Factor in Ru$^{3+}$}

\author{Colin L. Sarkis}
\email{sarkiscl@ornl.gov}
\affiliation{Neutron Scattering Division, Oak Ridge National Laboratory, Oak Ridge, Tennessee, 37831, USA}
\affiliation{Quantum Science Center, Oak Ridge National Laboratory, Oak Ridge, Tennessee, 37831, USA}
\author{John W. Villanova}
\affiliation{Center For Nanophase Materials Sciences, Oak Ridge National Laboratory, Oak Ridge, Tennessee 37831, USA}
\author{Casey Eichstaedt}
\affiliation{Department of Physics and Astronomy, The University of Tennessee, Knoxville, TN 37996, USA }
\affiliation{Institute for Advanced Materials \& Manufacturing, The University of Tennessee, Knoxville, TN 37920, USA}
\affiliation{National Renewable Energy Laboratory, Golden, CO 80401, USA}
\author{Adolfo G. Eguiluz}
\affiliation{Department of Physics and Astronomy, The University of Tennessee, Knoxville, TN 37996, USA }
\affiliation{Institute for Advanced Materials \& Manufacturing, The University of Tennessee, Knoxville, TN 37920, USA}
\author{Jaime A. Fernandez-Baca}
\affiliation{Neutron Scattering Division, Oak Ridge National Laboratory, Oak Ridge, Tennessee, 37831, USA}
\author{Masaaki Matsuda}
\affiliation{Neutron Scattering Division, Oak Ridge National Laboratory, Oak Ridge, Tennessee, 37831, USA}
\author{Jiaqiang Yan }
\affiliation{Materials Science and Technology Division, Oak Ridge National Laboratory, Oak Ridge, Tennessee 37831, USA}
\affiliation{Quantum Science Center, Oak Ridge National Laboratory, Oak Ridge, Tennessee, 37831, USA}
\author{Christian Balz}
\affiliation{ISIS Neutron and Muon Source, STFC Rutherford Appleton Laboratory, Didcot OX11 0QX, United Kingdom}
\author{Arnab Banerjee}
\affiliation{Department of Physics and Astronomy, Purdue University, West Lafayette IN - 47907, USA}
\affiliation{Quantum Science Center, Oak Ridge National Laboratory, Oak Ridge, Tennessee, 37831, USA}
\author{D. Alan Tennant}
\affiliation{Department of Physics and Astronomy, The University of Tennessee, Knoxville, TN 37996, USA }
\affiliation{Quantum Science Center, Oak Ridge National Laboratory, Oak Ridge, Tennessee, 37831, USA}
\affiliation{Shull Wollan Center - A Joint Institute for Neutron Sciences, Oak Ridge National Laboratory, Oak Ridge, Tennessee 37831, USA}
\author{Tom Berlijn}
\affiliation{Center For Nanophase Materials Sciences, Oak Ridge National Laboratory, Oak Ridge, Tennessee 37831, USA}
\affiliation{Quantum Science Center, Oak Ridge National Laboratory, Oak Ridge, Tennessee, 37831, USA}
\author{Stephen E. Nagler}
\email{naglerse@ornl.gov}
\affiliation{Neutron Scattering Division, Oak Ridge National Laboratory, Oak Ridge, Tennessee, 37831, USA}
\affiliation{Quantum Science Center, Oak Ridge National Laboratory, Oak Ridge, Tennessee, 37831, USA}
\affiliation{Department of Physics and Astronomy, The University of Tennessee, Knoxville, TN 37996, USA }

\date{\today}
%
%
%
%
\begin{abstract}
The Mott insulator \rucl\ has generated great interest in the community due to its possible field-induced Kitaev quantum spin liquid state. Despite enormous effort spent trying to obtain the form of the low energy Hamiltonian, there is currently no agreed upon set of parameters which is able to explain all of the data. A key piece of missing information lies in the determination of the magnetic form factor of Ru$^{3+}$, particularly for a true quantitative treatment of inelastic neutron scattering data. Here we present the experimentally derived magnetic form factor of Ru$^{3+}$ in the low spin 4$d^5$ state using polarized neutron diffraction within the paramagnetic regime on high quality single crystals of \rucl. We observe strong evidence of an anisotropic form factor, expected of the spin-orbit coupled $j_{\textrm{eff}} = \frac{1}{2}$ ground state. We model the static magnetization density in increasing complexity from simple isotropic cases, to a multipolar expansion, and finally \emph{ab initio} calculations of the generalized $j_{\textrm{eff}} = \frac{1}{2}$ ground state.  Comparison of both single ion models and inclusion of Cl$^-$ anions support the presence of hybridization of Ru$^{3+}$ with the surrounding Cl$^{-}$ ligands.
\end{abstract}

\maketitle

\section{Introduction}
Kitaev's exactly solvable model on a honeycomb lattice with bond-dependent nearest-neighbor interactions has generated enormous interest due to its quantum spin liquid ground state, where the excitations of the model can be described by itinerant Majorana fermions and static $\mathrm{Z}_2$ fluxes~\cite{kitaev2006anyons}. An exciting possible extension of the model that Kitaev realized was its application as a framework for fault tolerant quantum computation~\cite{kitaev2006anyons,kitaev2003fault,nayak2008non}, where application of a magnetic field in the gapless phase generates non-abelian anyons. Finding materials which can realize the Kitaev model is therefore of great interest both in the search for a quantum spin liquid, and as a route to the development of fault tolerant quantum computation.

The search for Kitaev candidate materials has mostly centered around 4$d^5$ and 5$d^5$ transition metals after it was realized that the bond-dependent interactions of the Kitaev model could be realized through the combination of spin-orbit coupling (SOC) and edge-sharing octahedral ligand fields~\cite{jackeli2009mott,chaloupka2010kitaev}. Among possible materials which exhibit these characteristics, \rucl\ is one of the most promising candidates. Though High-quality crystals show zigzag antiferromagnetic order below $T_{N}$ = 7~K, it was shown that application of an external magnetic field within the honeycomb plane (perpendicular to a Ru-Ru bond) can destroy magnetic order for fields above 7.3~T~\cite{banerjee2018excitations}. Signatures of fractionalized excitations in zero applied field, as well as at intermediate magnetic fields where magnetic order is suppressed, have been observed in inelastic neutron scattering (INS)~\cite{banerjee2016proximate,do2017majorana,banerjee2017neutron}, nuclear magnetic resonance~\cite{baek2017evidence}, terahertz spectroscopy~\cite{little2017antiferromagnetic}, specific heat measurements~\cite{widmann2019thermodynamic,tanaka2022thermodynamic}, and Raman spectroscopy~\cite{mai2019polarization,wulferding2020magnon}. 

Most intriguingly, quantized fractional plateaus in thermal hall measurements have been reported~\cite{kasahara2018majorana}. The values of the plateaus match the expected value for Majorana fermion edge currents associated with the Kitaev model, lending strong support to \rucl\ being a candidate Kitaev material. However the challenging nature of these measurements has led to much debate regarding the nature of these plateaus, with studies reporting no observation of fractional quantized plateaus~\cite{czajka2021oscillations}, sample dependence of the thermal conductivity and Hall plateaus~\cite{yamashita2020sample,zhang2023sample}, and even reports of anomalous fractional quantized plateaus~\cite{yokoi2021half}.

Despite the enormous effort characterizing \rucl, there is still no low energy Hamiltonian which can reproduce all experimental data. While \rucl\ has been found to host dominant Kitaev exchange, other terms such as isotropic Heisenberg exchange, off-diagonal exchange, and further neighbor interactions are needed to stabilize the experimentally observed zigzag antiferromagnetic order at low temperatures~\cite{banerjee2016proximate,winter2017models,suzuki2018effective,eichstaedt2019deriving}. One problem preventing further determination of the Hamiltonian, lies in accurately describing the magnetic form factor of Ru$^{3+}$. Current efforts to include the magnetic form factor in calculations have relied on simple approximations, with some papers utilizing Dirac-Slater calculations~\cite{cromer1964alamos,banerjee2017neutron}, while others have used the form factor of Fe$^{3+}$~\cite{ritter2016zigzag}. Importantly, the $j_{\textrm{eff}} =\frac{1}{2}$ ground state of magnetic $4d^5$ Ru ions is expected to be highly anisotropic, which may lead to significant deviations in the form factor from these simple cases. The true nature of the form factor of Ru$^{3+}$ can provide insight into the bonding nature between the Ru and Cl atoms and provide another experimental constraint to allow for a more accurate determination of the generalized Kitaev Hamiltonian in \rucl, in particular by providing quantitative fits to scattering data. 

In this work, we study the local electronic state of Ru$^{3+}$ in the low spin state by directly measuring the magnetic structure factor of Ru$^{3+}$ through polarized neutron diffraction (PND). We attempt to model the magnetic form factor using several separate models: (i) a spherical single ion form factor similar to that used in Ref. \onlinecite{banerjee2017neutron}, (ii) a spherical ionic form with moments allowed on neighboring Cl$^{-}$ ligands, (iii) an anisotropic single ion multipole expansion, and (iv) \emph{ab~initio}  calculations of the generalized $j_{\textrm{eff}} =\frac{1}{2}$ ground state using Wannier functions in order to accurately account for $d-p$ orbital mixing with the Cl$^-$ ions. Our results show magnetic form factor of Ru$^{3+}$ in the low-spin state in \rucl\ is anisotropic.

\section{Experimental Methods}
2~g high quality single crystals of \rucl\ were grown by sublimating purified powder of \rucl, following Ref. \onlinecite{PhysRevMaterials.7.013401}. Polarized neutron diffraction was performed at PTAX (HB-1), the Polarized Neutron Triple-Axis instrument at the High Flux Isotope Reactor in Oak Ridge National Laboratory. The samples were mounted in a vertical field asymmetric cryomagnet. All data was collected at 80~K, well into the paramagnetic phase of \rucl, but below the 1$^{st}$ order structural phase transition into the proposed $R \bar{3}$ structure. For both samples a vertical field was applied out of the nominal scattering plane. Sample ``A'' had an 8~T field oriented perpendicular to a Ru-Ru bond such that the $(h$~$0$~$l)$ plane was contained within the scattering plane. For sample ``B'', a magnetic field of 6~T was oriented parallel to a Ru-Ru bond, such that the scattering plane contained the $(h$~$h$~$l)$ plane. The instrument was run in half-polarized mode for the majority of the data collection, with the incident beam polarized. Polarized neutrons with incident energies of $E_i$ = 13.5~meV and $E_i$ = 30.5~meV were selected using a Heusler alloy monochromator. PG filters were installed to reduce $\lambda$/2 contamination and a PG analyzer was chosen on the scattered side to minimize background signals. Open collimation settings of 48-80-60-240 were used to ensure a full integration of Bragg peaks. Data were collected using three-point scans with flippers on and off, with typical count times of 10~min per point. In order to prevent sample movement due to magnetic torque in-field, the samples were secured in a large aluminum sample mount, which provided a non-magnetic determination of the beam polarization and flipper efficiencies (Appendix \ref{sec:Neutron}).
\begin{figure*}[!htb]
    \centering
    \includegraphics[width=\textwidth]{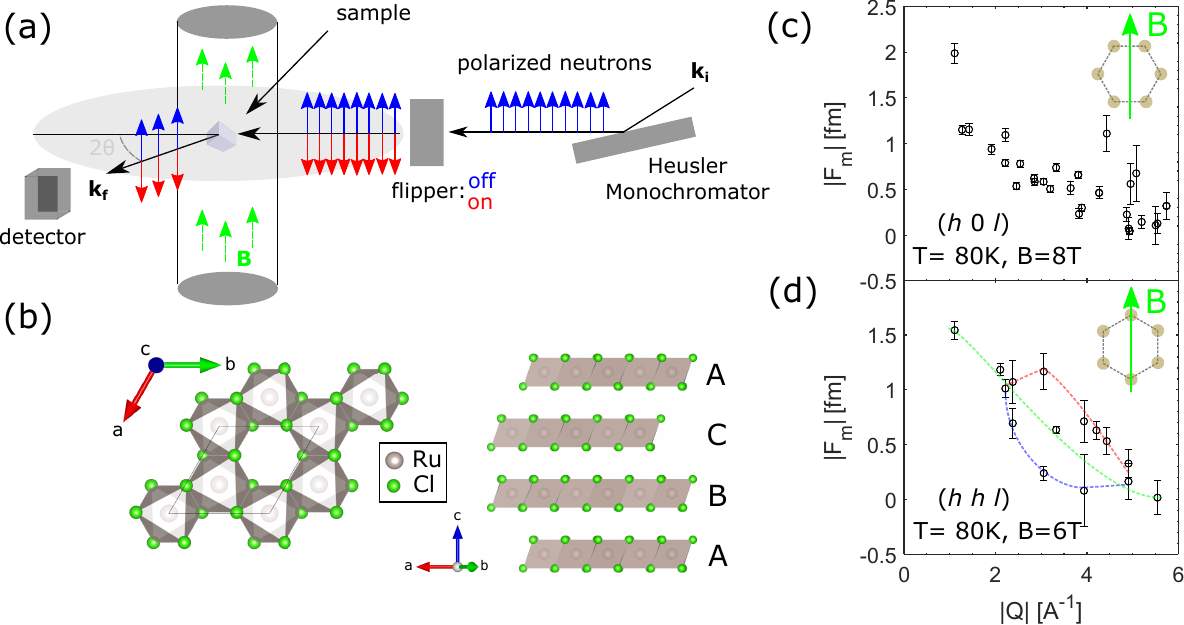}
    \caption{(a) Schematic of half-polarized neutron diffraction. Incoming neutrons are polarized to be parallel to the applied external field magnetic field at the sample position. An RF flipper can be used to flip the neutron spin to be oppositely-aligned with the magnetic field. Scattered elastic neutrons are counted in a polarization-insensitive detector. (b)  Three-layer stacking $R \bar{3}$ crystal structure used for all calculations in the work, from Ref. \onlinecite{mu2021role}. (c) Absolute value of magnetic structure factors taken at 80~K in the ($h$~0~$l$) plane. Inset shows field direction of 8~T field applied perpendicular to a Ru-Ru bond. (d) Absolute value of magnetic structure factors taken at 80~K in the ($h$~$h$~$l$) plane. Colored dashed lines are guides to the eye showing the three main branches of reflections: red corresponds to ($h$~$h$~$l$) type reflections with both $h,l$ $>$ 0, green corresponds to reflections where either $h$ or $l$ are equal to zero, and blue corresponds to ($h$~$h$~$\bar{l}$) type reflections where both $h,l$ $>$ 0. Inset shows direction of 6~T field applied parallel to a Ru-Ru bond.}
    \label{fig:PND}
\end{figure*}

\section{Results \& Discussion}

Polarized neutron studies have been extensively used to extract the form factors of many complex magnetic ions~\cite{nathans1959use,moon1966magnetic,moon1972distribution,schweizer1980polarised,brown2010magnetization,nandi2013magnetization,jeong2020magnetization}. In the paramagnetic regime, application of a magnetic field will partially polarize the sample along the direction of the field. Neutrons with spins polarized along the field ($I_{+}$) and opposite to the field ($I_-$) contribute to the elastic scattered intensity of a crystalline Bragg peak through both nuclear and magnetic scattering. By measuring with the system in a half-polarized geometry (see Fig. \ref{fig:PND}a), where the initial beam is polarized and the scattered beam is insensitive to the polarization, one can look at the ratios of these intensities. The ratio of the spin up ($I_{+}$) and spin down ($I_-$) intensities leads to a quantity known as the flipping ratio ($R$), which for a centrosymmetric crystal (real-valued structure factors) takes the form

\begin{equation}
    R = \frac{I_+}{I_-} = \frac{F_{N}^2 + 2psin^2(\alpha)F_{N}F_{M} + sin^2(\alpha)F_{M}^2}{F_{N}^2 - 2p\epsilon sin^2(\alpha)F_{N}F_{M} + sin^2(\alpha)F_{M}^2}
    \label{eq:flip_ratio}
\end{equation}
where $F_{M}$ and $F_{N}$ are the magnetic and nuclear structure factors respectively, $p$ is the initial beam polarization efficiency, $\epsilon$ is the flipper efficiency, and the angle $\alpha$ is the angle between the scattering plane and static component of the magnetization. 

In order to determine the magnetic structure factor, $F_M$ (from which one can determine the magnetic form factor), Eq.~\ref{eq:flip_ratio} is explicitly dependent on one's choice of nuclear unitcell, which is determined separately and taken to be a known parameter. For \rucl, this is a particularly important distinction. While the room temperature structure has been largely confirmed by the community to be $C 2/m$, \rucl\ has been shown by multiple groups to host a strongly $1^{st}$ order phase transition below $T \sim 125~K$ in many samples~~\cite{park2016emergence,glamazda2017relation,lampen2018anisotropic}. The structure of the low temperature phase has so far eluded an exact determination, hindered both by the existence of structural domains~\cite{park2016emergence} and the material's sensitivity toward developing stacking faults upon handling~\cite{cao2016low}. Groups have reported low temperature structures such as $P 3_1 1 2$~\cite{banerjee2016proximate,ziatdinov2016atomic}, $R \bar{3}$~\cite{park2016emergence,reschke2018sub,mu2021role}, and $C 2/m$~\cite{cao2016low,little2017antiferromagnetic}. For all analysis shown in this work we use the recently reported low temperature structure with the lowest reported R factor found in Ref. \cite{mu2021role} who found an $R \bar{3}$ structure ($a = 5.973\AA$, $c = 17.0025~\AA$ in hexagonal setting) shown in Fig. \ref{fig:PND}b.

We focus on small $|\vec{Q}|$ reflections which are most sensitive to the magnetization and measured the flipping ratios for 28 unique reflections within the $(h$~$0$~$l)$ plane and 19 unique reflections within the $(h$~$h$~$l)$ plane. Along with the standard obverse setting reflection conditions of $R\bar{3}$ (-$h$+$l$ = 3$n$ for $k$=0), in our experiment we observed scattering belonging to a second domain consistent with a reverse setting ($h$+$l$ = 3$n$ for $k$=0). This type of twinning is consistent with that found by Ref. \onlinecite{park2016emergence} and can be related to the primary domain via a $c$-axis mirror. 

For $(h$~$0$~$l)$ measurements, by looking at multiple opposite domain reflections [e.g. ($\bar{2}~0~4$) vs ($2~0~4$)], we find that both domains were roughly equally populated with similar intensities and identical flipping ratios within measured uncertainty. Reflections which are allowed by both domains (e.g. $(3n$~$0$~$3n')$) mix \emph{equivalent} reflections and thus do not affect determination of $F_{m}$ (Appendix \ref{sec:Neutron}). Reflections taken in $(h$~$h$~$l)$ plane where both $h$ and $l$ are non-zero mix \emph{non-equivalent} reflections, such as $(1$~$1$~$3)$ and $(1$~$1$~$\bar{3})$. We can incorporate the effect on the intensities and flipping ratios for two domains as the sum of two independent reflections.
\begin{equation}
    R^{hhl} = \frac{a_p I^{hhl}_{+} + (1-a_p )\Tilde{I}^{hhl}_{+}}{a_p I^{hhl}_{-} + (1-a_p )\Tilde{I}^{hhl}_{-}}
\label{eq:domains}
\end{equation}
where $a_p$ represents the primary domain population and $I^{hhl}_{+,-}$, $\Tilde{I}^{hhl}_{+,-}$ correspond to the intensity from the primary and twinned domain respectively, according to Eq. \ref{eq:flip_ratio}. Surprisingly the best fit domain population was nearly monodomain, with $a_p$ = 0.95(5) which may be a result of careful annealing protocol used across the structural transition~(Appendix \ref{sec:Neutron}). Values of $F_{m}$ for these reflections were then able to be extracted for measurements with corresponding paired flipping ratios. For all data reported in this manuscript, we report only values corresponding to the primary structural domain.

\begin{table}[!]
\caption{Experimental data of flipping ratios and corresponding calculated structure factors for the $(h$~$0$~$l)$ plane in \rucl. All data is taken at 80~K and with an 8~T applied field along a $\{1$~$\bar{1}$~$0\}$ equivalent direction.}
\label{tab:data_h0l}
\begin{tabular}{cccc}

\hline
$(h~k~l)$~~~        & $R$~~~         & $F_{N}$ [fm]~~~         & $F_{M}$ [fm]~~~               \\ 
\hline
$(3~0~0)$~~~        & 1.008(1)~~~   & 210.874~~~              & 0.64(6)~~~                   \\
$(1~0~1)$~~~        & 1.147(25)~~~   & 28.109~~~               & 1.16(5)~~~                   \\
$(\bar{2}~0~1)$~~~  & 1.35(15)~~~    & 5.962~~~                & 0.54(4)~~~                   \\
$(4~0~1)$~~~        & 1.016(5)~~~    & 46.390~~~               & 0.23(7)~~~                   \\
$(\bar{1}~0~2)$~~~  & 1.158(2)~~~    & -26.012~~~              & -1.16(6)~~~                    \\
$(2~0~2)$~~~        & 1.128(5)~~~    & -10.707~~~              & -0.79(4)~~~                    \\
$(\bar{4}~0~2)$~~~  & 1.01(1)~~~    & -39.280~~~              & -0.1(1)~~~                     \\
$(0~0~3)$~~~        & 1.118(2)~~~    & -58.511~~~              & -2.0(1)~~~                     \\
$(3~0~3)$~~~        & 1.038(1)~~~    & -58.189~~~              & -0.67(4)~~~                    \\
$(1~0~4)$~~~        & 1.196(5)~~~    & -17.498~~~              & -0.94(6)~~~                    \\
$(\bar{2}~0~4)$~~~  & 1.075(4)~~~    & -28.607~~~              & -0.62(4)~~~                    \\
$(4~0~4)$~~~        & 1.3(1)~~~     & -7.693~~~               & -0.7(3)~~~                     \\
$(\bar{1}~0~5)$~~~  & 1.193(5)~~~    & 15.004~~~               & 0.79(4)~~~                   \\
$(2~0~5)$~~~        & 1.059(2)~~~    & 34.251~~~               & 0.59(3)~~~                   \\
$(\bar{4}~0~5)$~~~  & 1.9(2)~~~      & -0.763~~~               & -0.15(6)~~~                    \\
$(0~0~6)$~~~        & 0.972(1)~~~    & -127.112~~~             & 1.15(7)~~~                   \\
$(3~0~6)$~~~        & 0.987(15)~~~   & -123.480~~~             & 0.47(65)~~~                   \\
$(1~0~7)$~~~        & 1.157(6)~~~    & 13.391~~~               & 0.59(4)~~~                   \\
$(4~0~7)$~~~        & 0.95(1)~~~      & -6.748~~~               & 0.1(2)~~~                    \\
$(\bar{1}~0~8)$~~~  & 1.119(6)~~~    & -15.015~~~              & -0.51(3)~~~                    \\
$(2~0~8)$~~~        & 1.023(5)~~~    & -33.967~~~              & -0.23(4)~~~                    \\
$(0~0~9)$~~~        & 0.67(2)~~~     & 6.226 ~~~               & -0.75(6)~~~                    \\
$(3~0~9)$~~~        & 0.97(25)~~~    & 5.216~~~                & -0.04(4)~~~                    \\
$(1~0~10)$~~~       & 1.044(5)~~~    & -23.2235~~~             & -0.30(4)~~~                    \\
$(0~0~12)$~~~       & 1.018(3)~~~    & 202.300~~~              & 1.2(2)~~~                    \\
$(3~0~12)$~~~       & 1.005(2)~~~    & 198.860~~~              & 0.3(15)~~~                   \\
$(1~0~13)$~~~       & 1.06(2)~~~     & 29.193~~~               & 0.6(2)~~~                    \\
$(0~0~15)$~~~       & 1.003(3)~~~    & -120.924~~~             & -0.1(1)~~~                     \\\hline
\end{tabular}

\end{table}

\begin{table}[!]
\caption{Experimental data of flipping ratios and corresponding calculated structure factors for the $(h$~$h$~$l)$ plane in \rucl. All data is taken at 80~K and with a 6~T applied field along a $\{1$~$0$~$0\}$ equivalent direction. For all reflections with both $h \neq$ 0 and $l \neq$ 0, the flipping ratios contain components from two nonequivalent reflections due to a $c$-axis twin in the measured crystal. All $F_M$ reported contain only the extracted value pertaining to the primary domain.}
\label{tab:data_hhl}
\begin{tabular}{cccc}

\hline
$(h~k~l)$~~~        & $R$~~~         & $F_{N}$ [fm]~~~         & $F_{M}$ [fm]~~~              \\ 
\hline
$(1~1~0)$~~~        & 0.914(2)~~~    & -43.425~~~              & 1.18(6)~~~                  \\
$(2~2~0)$~~~        & 0.952(6)~~~    & -41.318~~~              & 0.63(8)~~~                  \\
$(0~0~3)$~~~        & 1.086(1)~~~    & -58.511~~~              & -1.54(8)~~~                   \\
$(1~1~3)$~~~        & 1.013(15)~~~    & -181.607~~~             & -1.1(2)~~~                   \\
$(1~1~\bar{3})$~~~  & 0.9850(9)~~~   & 113.467~~~              & -0.70(4)~~~                   \\
$(0~0~6)$~~~        & 0.978(1)~~~    & -127.112~~~             & 1.01(8)~~~                  \\
$(1~1~6)$~~~        & 1.021(25)~~~    & 154.202~~~              & 1.2(2)~~~                   \\
$(1~1~\bar{6})$~~~  & 1.006(1)~~~    & 98.295~~~               & 0.24(5)~~~                  \\
$(0~0~9)$~~~        & 0.71(2)~~~     & 6.226~~~                & -0.64(4)~~~                   \\
$(1~1~9)$~~~        & 0.977(3)~~~    & 76.023~~~               & -0.71(9)~~~                   \\
$(1~1~\bar{9})$~~~  & 0.999(1)~~~    & -208.459~~~             & 0.1(3)~~~                   \\
$(0~0~12)$~~~       & 1.008(2)~~~    & 202.300~~~              & 0.5(1)~~~                  \\
$(1~1~12)$~~~       & 0.990(4)~~~    & -92.238~~~              & 0.3(1)~~~                   \\
$(1~1~\bar{12})$~~~ & 1.01(2)~~~     & 17.569~~~               & 0.2(1)~~~                  \\
$(0~0~15)$~~~       & 1.000(4)~~~    & -120.924~~~             & 0.0(15)~~~                   \\\hline
\end{tabular}

\end{table}

\begin{figure*}[!htb]
    \centering
    \includegraphics[width=\textwidth]{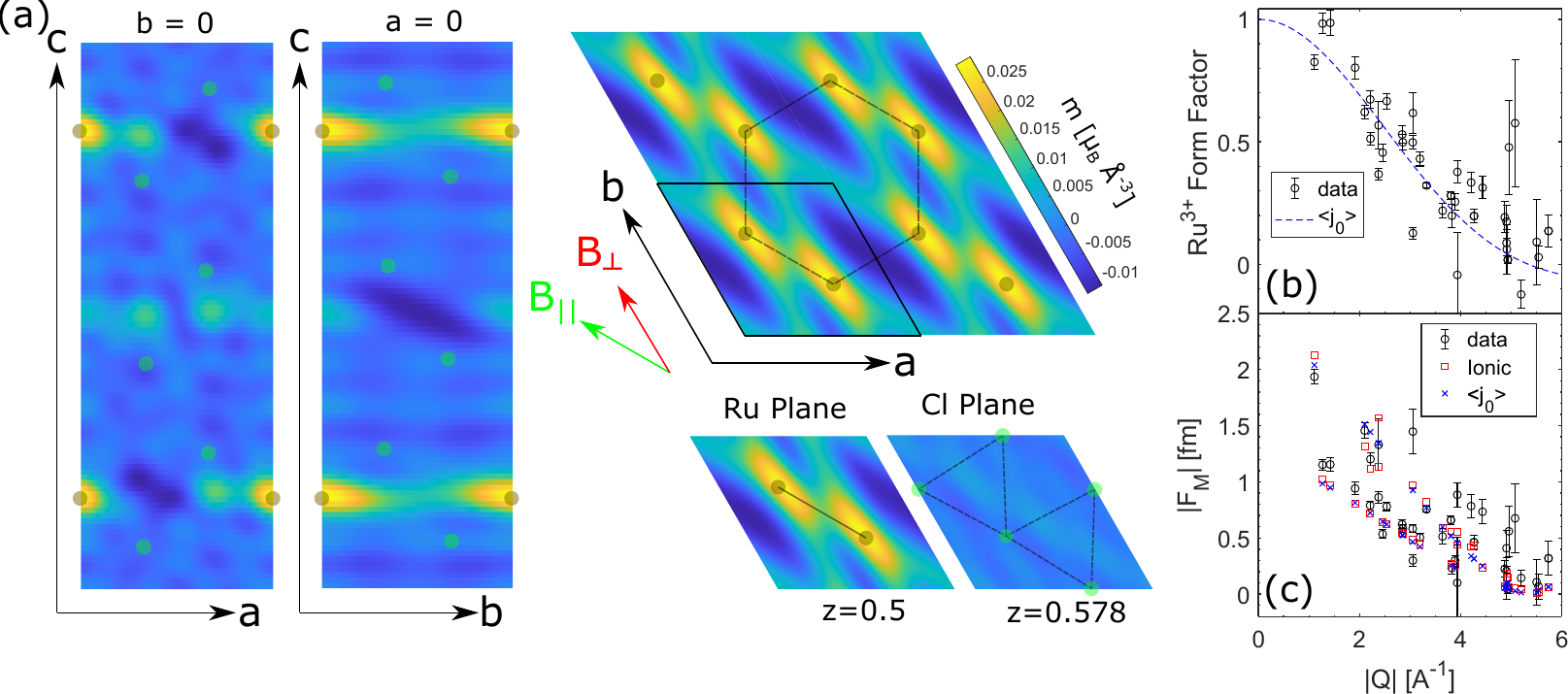}
    \caption{(a) Direct space magnetization density, m(r), produced directly through Eq. \ref{eq:mr} using data from Tables \ref{tab:data_h0l} and \ref{tab:data_hhl}. Brown and green circles denote Ru and Cl ions respectively and B$_\perp$ and B$_{||}$ show the defined magnetic field directions used for both experiments. The magnetization shows a weak induced moment centered on Ru ions which form into honeycomb layers. Observation of magnetization near surrounding Cl ions is hindered by a large amount of Fourier noise. (b) Normalized single-ion Ru$^{3+}$ form factor data extracted following Eq. \ref{eq:form_factor}. The dashed line shows best fit lowest order spherical approximation ($\braket{j_{0}}$). The data was normalized using a best-fit induced moment of $\mu = 0.145\mu_B/Ru$ such that the presented data is scaled to unity at $|\vec{Q}|$ = 0. The form factor shows clear signs of anisotropy away from a simple decreasing trend in $|\vec{Q}|$ that are clearly not accounted for without angular dependence. (c) Experimentally observed magnetic structure factor values compared to an ionic spherical model containing contributions from Cl$^{-}$ ions (Eq. \ref{eq:form_factor_ionic}) as well as values for the single-ion spherical model shown in (b).}
    \label{fig:isotropic}
\end{figure*}

Values of the experimentally measured flipping ratios and corresponding magnetic structure factor extracted using Eq. \ref{eq:flip_ratio} and Eq. \ref{eq:domains} are presented in Tables \ref{tab:data_h0l} and \ref{tab:data_hhl}. While Eq. \ref{eq:flip_ratio} gives two roots for each reflection, we find that only one root remains physical for each measured reflection. This allowed for an unambiguous measure of $F_{M}$. We note that since the extracted magnetic structure factor is phase sensitive, $F_{M}$ is not a definite positive value for all $\vec{Q}$. Fig. \ref{fig:PND}c-d show the absolute value of the derived magnetic structure factors as a function of $|\vec{Q}|$ for both orientations. While both data sets show similar overall decreasing trends for increasing momentum transfer, the ($h$~$0$~$l$) data are mostly constrained to a single trend while the ($h$~$h$~$l$) data appear to split into three distinct branches, highlighted by the different colored dashed lines. These branches correspond to the following reflection types: red show ($h$~$h$~$l$) type reflections with both $h$ $>$ 0, $l$ $>$ 0, green show reflections where either $h$ or $l$ are zero, and blue show ($h$~$h$~$\bar{l}$) type reflections with with $h$ and $l$ finite and opposite in sign. 

In order to obtain the direct-space scalar magnetization density, $m(\vec{r})$, one can simply take a direct Fourier transform of the measured magnetic structure factors. Since the magnetization density has the periodicity of the crystal lattice, one can define
\begin{equation}
    p_0m(\vec{r}) =\frac{1}{v_0} \sum_G F_M(\vec{G})e^{i\vec{G} \cdot \vec{r}}
    \label{eq:mr}
\end{equation}
where the magnetic scattering length $p_0$ is a well-known constant ($\frac{\gamma r_{0}}{2\mu_{B}} = 2.695 \frac{\text{fm}}{\mu_{B}}$, with $\gamma$ the ratio of the neutron magnetic dipole moment to the nuclear magneton ($\gamma>0$), and $r_0$ the classical electron radius), $v_0$ is the volume of the unitcell and the sum runs over the complete set of reciprocal lattice vectors $\vec{G}$. Fig. \ref{fig:isotropic}a shows $m(\vec{r})$ produced via a direct Fourier transform of the data in Tables \ref{tab:data_h0l} and \ref{tab:data_hhl}. A lack of Fourier components, both from a lack of data perpendicular to the measured scattering planes as well as from finite measured $|\vec{Q}|$ within the scattering planes, introduce a large amount of structured Fourier noise to a direct Fourier transform. Despite this, a few things are clear: (i) the magnetization density shows an induced moment centered on Ru$^{3+}$ ions, (ii) the magnetization density shows elongation along $b$ which could be related to the low temperature preference for zigzag order, but is most likely due to larger noise as $b$ corresponds to the direction in reciprocal space with the least experimental coverage, (iii) no obvious moment beyond background can be made out on neighboring Cl$^{-}$ ions which constrain the relative moment on Cl$^{-}$  to be roughly $<$ 25\% the total moment. 

In order to more reliably extract an accurate depiction of the magnetization density, one can fit the magnetic structure factor directly using a known model. To separate the normalized magnetic form factor from the measured magnetic structure factors, one can often model the magnetization density using a single ion approach, where the magnetization density is constrained to only involve the magnetic ion species and all interactions with surrounding ions are ignored, or with a more inclusive multi-ion model which features the addition of the surrounding ligand environment beyond the single-ion level.


\subsection{Isotropic Models}

For an isolated ion in free-space, the magnetization density due to unpaired electrons is spherical and can defined by a radial wavefunction, $R(r)$. The magnetic form factor is then described by radial integrals $\braket{j_l(Q)}$ = $\int R(r)^2j_{l}(Qr)4\pi dr$, where $j_l(Qr)$ are spherical Bessel functions of order $l$. For all analysis in this work, we use radial functions of the form described in Appendix \ref{sec:multipole}. Since the magnetic form factor of transition metals decay quickly with $|\vec{Q}|$, it is often sufficient to keep only the lowest order terms. When embedded in solids the magnetization density can become anisotropic. Despite this a large number of spherical form factors remain successful in describing the experimental data. An extensive list of these form factors have been tabulated by Brown~\cite{Brown2006magnetic}, though we note that Ru$^{3+}$ is not among those tabulated. In order to compare our derived form factor to a simple lowest order spherical form, we compare Dirac-Slater calculations of the Ru$^{3+}$ form factor using the contributions of the unfilled orbitals of the x-ray form factor, which were produced by Cromer and Waber in Ref. \onlinecite{cromer1964alamos}. When we compare fits to Dirac-Slater calculations we find within our experimental limit of $|\vec{Q}|$ it can be well reproduced by the lowest order radial integral, $\braket{j_0}$, with a radial parameter of $\xi$ = 6.4.

For a single magnetic species, the magnetic structure factor can be written as
\begin{equation}
    F_{M}(\vec{Q}) = p_0  \mu f(\vec{Q})\sum_{j} e^{i\vec{Q} \cdot \vec{r}_{j}}e^{-W_j} 
    \label{eq:form_factor}
\end{equation}
where $\mu$ represents the magnetic moment in units of $\mu_{B}$ (Bohr magnetons) per ion for a symmetry equivalent atom, $f(\vec{Q})$ is the normalized form factor, $e^{-W_j}$ is the Debye-Waller Factor, and the sum runs over all magnetic ions in the unitcell.

The measured normalized (single ion) magnetic form factor taken using Eq. \ref{eq:form_factor} is shown in Fig. \ref{fig:isotropic}b. Here we scaled the experimental data with a best fit induced moment of $\mu$ = 0.141(5)$\mu_B$/Ru, such that the magnetic form factor is normalized to be unity at $\vec{Q}$ = 0. For $(h$~$h$~$l)$ data taken with an applied field of 6~T, we compare common $(0$~$0$~$l)$ reflections in order to provide a relative induced moment such that we can combine both datasets (Appendix \ref{sec:Neutron}). While most of the data show a similar decreasing trend in $|\vec{Q}|$ clear anisotropy is observed for some directions. Since $f(\vec{Q})$ shows clear directional dependence rather than just as a function of $|\vec{Q}|$, we have strong evidence that the form factor contains angular dependence which can \emph{not} be explained using spherical models, even with the inclusion of an orbital moment. While describing anisotropic data using such a model is prone to error, we find the best-fit lowest order spherical model described by $\braket{j_0}$ is able to describe much of the $(h$~$0$~$l)$ data well. We find fits of the radial distribution to the data are in good agreement to that calculated in Ref. \onlinecite{cromer1964alamos}, indicating that such a model was able to accurately describe the main radial extent of the form factor, if not the anisotropy.

 Eq. \ref{eq:form_factor} explicitly ignores hybridization effects by constraining magnetic form factor to only consider magnetic Ru$^{3+}$ ions. A common simple ionic approximation which can include hybridization effects is to extend Eq. \ref{eq:form_factor} to include a second sum over surrounding ligand ions. In this case we have
\begin{equation}
    F_{M}(\vec{Q}) = F^{Ru}_{M}(\vec{Q}) + F^{Cl}_{M}(\vec{Q})
    \label{eq:form_factor_ionic}
\end{equation}
where $F^{Ru}_{M}(\vec{Q})$ and $F^{Cl}_{M}(\vec{Q})$ corresponding to individual ionic single ion magnetic structure factors for each symmetry equivalent Ru$^{3+}$ and Cl$^-$ ions respectively. One can then vary the relative size of the moments on Ru and Cl sites as a free parameter under the constraint that the overall moment is fixed~\cite{moon1975magnetic,javorsky2001magnetization,kernavanois2003magnetization}. 

The addition of Cl$^-$ ions requires two independent form factors, which can double the parameters required to describe the data. Because of this, ionic models incorporating ligand atoms often assume simple spherical form factors for each atom. While this may not be an accurate model for anisotropic spin-orbit coupled moments, it can provide some insight into the bonding nature of \rucl. Fig. \ref{fig:isotropic}c shows best fit parameters corresponding to Eq. \ref{eq:form_factor_ionic}. In each case only the lowest order radial integrals ($\braket{j_0}$) were considered. Since each point in reciprocal space contains phase sensitive contributions from both Ru$^{3+}$ and Cl$^{-}$ ions, we compare the full magnetic structure factor rather than the magnetic form factors. While some of the data are clearly not accounted for, the addition of Cl tends to fit the data marginally better than modeling Ru alone, particularly at low $|\vec{Q}|$. The best fit parameters within this model find a sizeable portion of the total induced moment sit on Cl ion ($\sim 18\%$ the total moment). Best fit parameters corresponding to the two isotropic models described above can be found in Table \ref{tab:iso}. We can also compare the best fit moments to estimates of the induced moment calculated from Curie-Weiss fits to the intermediate temperature susceptibility, just below the structural transition but well above $T_N$ in comparison to bulk magnetization measurements described in~\cite{Li2021identification}. This leads to an induced moment of $\mu \sim 0.165(3)\mu_{B}$ for 80~K and an applied field of 8~T within the $ab$-plane, in good agreement with the total induced moment found with incorporation of the Cl atoms of $\sim$0.167$\mu_B$. 

\begin{table}[!]
\caption{Best fit parameters  using a lowest order single ion approximation ($\braket{j_0}$) and ionic approximation with moments allowed on Cl ions. Here $\zeta$ is a parameter describing the radial extent of the wavefunctions in units of $\AA^{-1}$ (Appendix \ref{sec:multipole}) while $\mu$ corresponds to the static induced magnet moment in units of $\mu_B$ such that the resulting form factor(s) is (are) normalized to unity at $|\vec{Q}|$=0. In the first column the model is compared to the Dirac Slater calculations in Ref. \onlinecite{cromer1964alamos} while the other two columns are fits to data extracted from Eq. \ref{eq:form_factor} and Eq.  \ref{eq:form_factor_ionic} respectively. }
\label{tab:iso}
\begin{tabular}{cccc}
\cline{2-4}
                                &~~Dirac-Slater~~       & ~~$\mu_{Ru}\braket{j_0}_{Ru}$~~ &~~~~~$\mu_{Ru}\braket{j_0}_{Ru} + \mu_{Cl}\braket{j_0}_{Cl}$~~~~~ \\ \cline{2-4}
$\zeta_{Ru}$~~     & 6.40(4)      & 6.45(2)             & 6.68(1)                            \\
$\zeta_{Cl}$~~     & -       & -                    & 2.91(2)                              \\
$\mu_{Ru}$~~       & -       & 0.145(4)             & 0.137(8)                           \\ 
$\mu_{Cl}$~~       & -       & -                    & 0.03(1)                           \\ \cline{2-4}
$\chi^2$           & 0.17       & 11.17                 & 10.02                               \\ \cline{2-4}
\end{tabular}
\end{table}

\subsection{Multipole Models}

\begin{figure*}
    \centering
    \includegraphics[width=\textwidth]{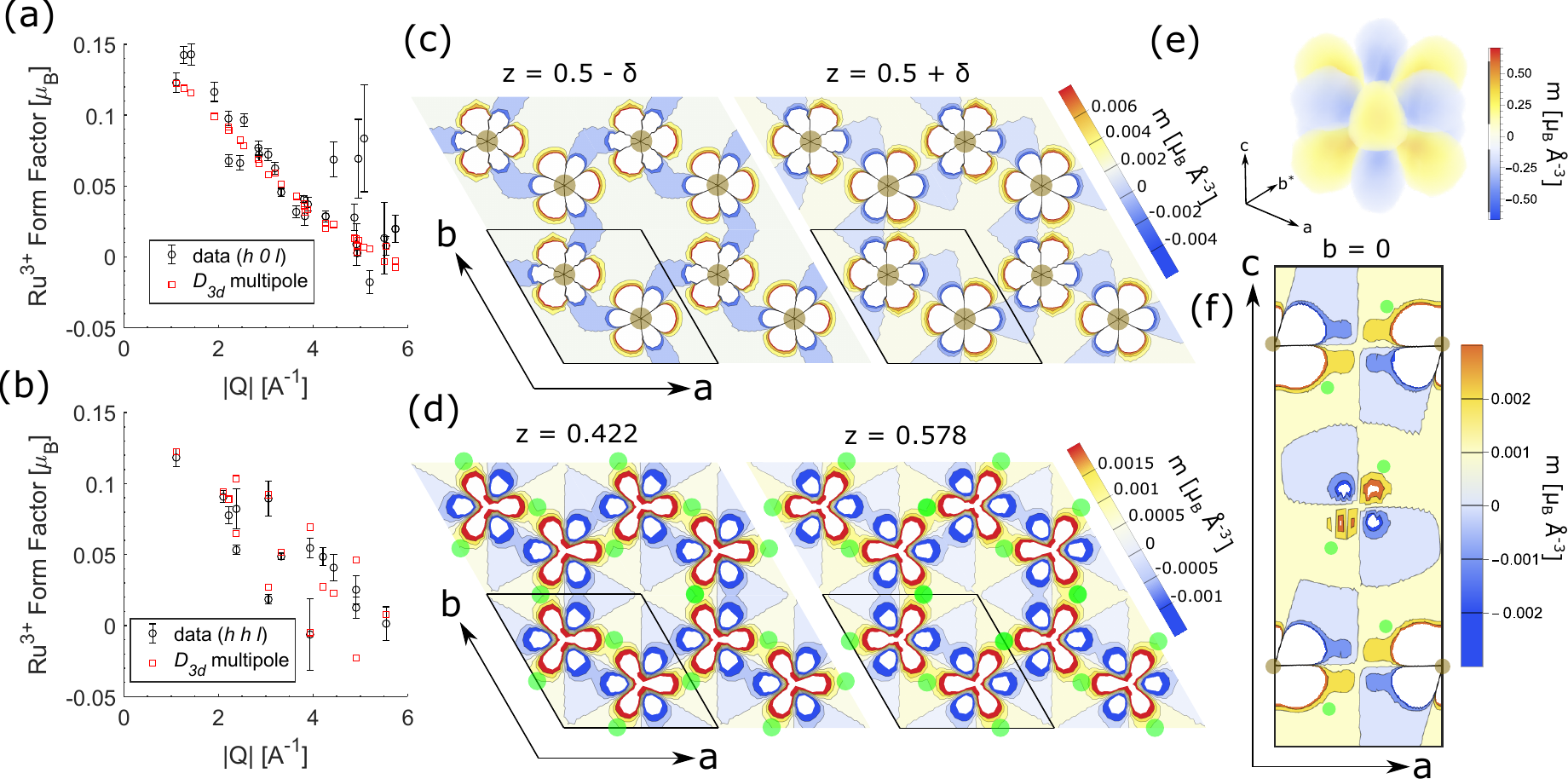}
    \caption{(a) Best fit $D_{3d}$ multipole model compared to the unnormalized single ion Ru$^{3+}$ form factor for data taken in the ($h~0~l$) plane. (b) Same best fit $D_{3d}$ multipole model compared to the unnormalized single ion Ru$^{3+}$ form factor for data taken in the ($h~h~l$) plane. Splitting is largely reproduced. (c) Contour plots of the direct space magnetization density of same $D_{3d}$ multipole model $\delta = 0.05 \AA$ above and below a nominal Ru honeycomb plane. (d) Contour plot of two nominal Cl planes above and below the Ru honeycomb layer. Brown and green spheres show the optimal positions of Ru and Cl respectively. Despite being a single-ion -model, the magnetization density shows a sizeable portion out to Cl positions. (e) Density plot of a single octahedra. The magnetization density shows two six-fold lobes with alternating sign extending above and below the honeycomb layer. (f) Contour plot of the magnetization density showing the out of plane three-layer stacking. }
    \label{fig:anisotropic}
\end{figure*}


Next, we turn to a single ion model which is appropriate for fitting complex systems of arbitrary symmetry. The multipole expansion has had success in systems with both spin and orbital contributions to the magnetization density~\cite{jeong2020magnetization}. For multipolar expansions, while the sum is infinite in principle the point group symmetry of the magnetic ion dictates which terms are non-zero~\cite{coppens1997x}. In this way fitting a multipolar expansion can provide a determination of the magnetization density without any assumptions about the ground state behavior. Details of this fit can be found in Appendix \ref{sec:multipole}. We restrict ourselves to only $l$ = even terms due to comparison with inversion symmetric $d$-orbitals. In this work we consider three different symmetries: (i) An $Oh_6$ symmetry due to a cubic octahedral arrangement of Cl$^-$ ions, (ii) a $D_{3d}$ trigonal distortion of the surrounding octahedra, and (iii) the reduced symmetry of the Ru site in the case of a $C_2$ monoclinic distortion. Best fit multipole parameters are listed in Table \ref{tab:multipole} for each case.

\begin{table}[!]
\caption{Best fit parameters of data to a general multipolar expansion (Eq. \ref{eq:multipole_fm}) up to $l$ = 4 for different symmetry conditions.}
\centering
\begin{tabular}{cccc}
\cline{2-4}
                        & $Oh_{6}$~~~           & ~~~$D_{3d}$~~~      & ~~~$C_{2}$    \\ \cline{2-4} 
$\zeta$~~~              & ~~~6.34(5)~~~         & ~~~6.826(5)~~~      & ~~~5.51(2)    \\ 
$\tilde{a}_{00}$~~~     & ~~~0.0384(4)~~~       & ~~~0.0388(4)~~~     &  ~~~0.041(7)  \\ 
$\tilde{a}_{20}$~~~     & ~~~0~~~               & ~~~-0.0012(5)~~~    &  ~~~0.001(9)   \\ 
$\tilde{a}^{c}_{22}$~~~ & ~~~0~~~               & ~~~0~~~             &  ~~~-0.066(5)  \\ 
$\tilde{a}^{s}_{22}$~~~ & ~~~0~~~               & ~~~0~~~             &  ~~~-0.0057(4)  \\ 
$\tilde{a}_{40}$~~~     & ~~~0.003(1)~~~        & ~~~0.0104(8)~~~     &  ~~~0.023(4)  \\ 
$\tilde{a}^{c}_{42}$~~~ & ~~~0~~~               & ~~~0~~~             &  ~~~-0.009(4)  \\ 
$\tilde{a}^{s}_{42}$~~~ & ~~~0~~~               & ~~~0~~~             &  ~~~0.029(4)  \\ 
$\tilde{a}^{c}_{43}$~~~ & ~~~0~~~               & ~~~-0.441(7)~~~     &  ~~~0           \\ 
$\tilde{a}^{s}_{43}$~~~ & ~~~0~~~               & ~~~0~~~             &  ~~~0         \\ 
$\tilde{a}^{c}_{44}$~~~ & ~~~0.012(1)~~~        & ~~~0~~~             &  ~~~-0.052(8)  \\ 
$\tilde{a}^{s}_{44}$~~~ & ~~~0~~~               & ~~~0~~~             &  ~~~-0.062(4) \\ \cline{2-4}
$\chi^2$~~~             & ~~~10.39~~~           & ~~~6.20~~~          &  ~~~7.31      \\\cline{2-4}
\end{tabular}
\label{tab:multipole}
\end{table}

In the cubic octahedral case, the spin axes naturally lie perpendicular to octahedra bonding axes (Fig. \ref{fig:CEF}), supporting isotropic Kitaev interactions. For a small perturbation due to an applied magnetic field in this case, one then expects the quantization axis becomes the component closest to the field direction. With addition of trigonal distortion or monoclinic distortion the natural $s_z$-axis becomes the direction of the compression or elongation corresponding to the crystallographic $c*$-axis. With this in mind, we modeled multipole fits for all three symmetries under both local conditions and find better fit for \emph{all} considered symmetries to the data assuming the local $s_z$-direction corresponds to the $c^*$-axis, consistent with a finite distortion of the Cl octahedral environment. Within a Kitaev perspective, aside from allowing other terms in the Hamiltonian, such as the Heisenberg and off-diagonal $\Gamma, \Gamma'$ terms, such a distortion may favor more of an anisotropic limit which has so far largely not been considered by the community. 

We find the best-fit to the data results from the trigonal distorted $D_{3d}$ symmetry, shown in Fig. \ref{fig:anisotropic}a-b. Further details of best-fits fits to other symmetries can be found in Appendix \ref{sec:multipole}. The model is able to reproduce the observed splitting at higher momentum transfers in the ($h~h~l)$ data, and reproduces the near isotropic behavior of the data in the ($h~0~l$) data. Similar to other fits, we find three data points at relatively high $|\vec{Q}|$ in the ($h~0~l$) plane associated with relatively poor counting statistics (hence large errors) which cannot be reproduced in our model. At low $|\vec{Q}|$ the model shows deviations away from the experimental data, implying that the multipole model is not accurately capturing the physics describing the form factor at low $|\vec{Q}|$. Similar to the spherical case, for a multipolar expansion at $|\vec{Q}|$ = 0 only one term ($\tilde{a}_{00}$) remains non-zero which can be compared to the induced magnetization. For all multipole fits, extracted values of $\tilde{a}_{00}$ = $\mu/\sqrt{4\pi}$ show a slightly smaller induced moment of $\mu \sim 0.138\mu_B$/Ru compared what was found for either spherical model. 

The magnetization density of the model is reproduced in Fig. \ref{fig:anisotropic}c-f. Dominated by a large $\tilde{Z}_{43}^c$ component, the magnetization density shows a double six-fold lobed structure with alternating sign to the magnetization density which is mirrored across the center of the honeycomb plane. Because the $\tilde{Z}_{43}^c$ features a node in the optimal honeycomb plane, the contour plot of the $ab$-plane shows a slight offset above the plane of $\delta = 0.05~\AA$ to better show the six-fold character of the magnetization density in the plane. Although treated as a single-ion model, the fit shows a fair spatial extent to the magnetization density, which also support the notion of hybridization in the data. This is best seen in the out of plane contour plot of the $ac$-plane

Another method which has been recently used to fit anisotropic magnetization densities is by including torodial dipoles as well as conventional axial dipoles, also known as anapoles~\cite{lovesey2015theory,lovesey2019direct,khalyavin2019anapole,lovesey2012magnetic}. In fact, anapoles recently been discussed in context of \rucl\ \cite{lovesey2022magnetization}, although that work used the monoclinic structure which has been generally associated with stacking fault-laden samples. Anapoles are expected to contribute to the elastic scattering intensities for any magnetic ions not located at inversion centers, such as the case for \rucl. While our approach does not identify the specific contribution of anapoles, we go beyond the toroidal dipole approximation. Instead, we perform \emph{ab initio} calculations of the magnetization density using Wannier functions which include all contributions present.

\subsection{Wannier Function Model}

\begin{figure}
    \centering
    \includegraphics[width=\columnwidth]{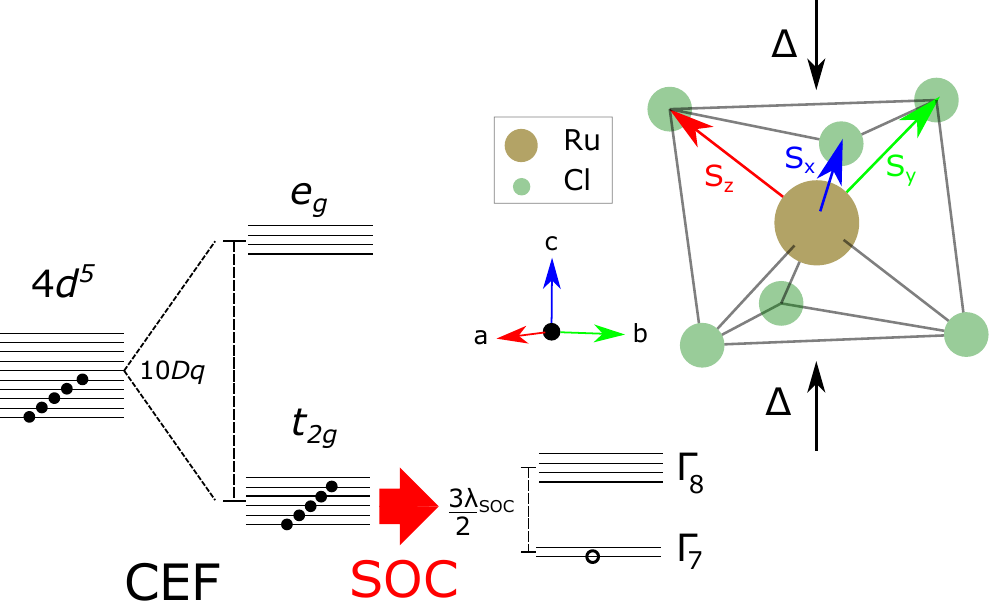}
    \caption{(Left) Energy diagram for Ru$^{3+}$ 4$d^5$ ions within the low-spin configuration of \rucl. The octahedral crystal field, leads to a large splitting between the $e_g$ and $t_{2g}$ bands (10$Dq \sim 2.4~eV$~\cite{suzuki2021proximate}. Sizeable SOC further splits the $t_{2g}$ into the doublet $\Gamma_7$ and quartet $\Gamma_8$. (Right) A single RuCl$_{6}$ octahedron, showing the relation of the spin axes ($S_{x},S_{y},S_{z}$) compared to the crystal coordinate system ($a,b,c$) in \rucl. For a perfectly cubic system, the spin axes lie along the Ru-Cl-Ru bonding axes with the unique z-axis chosen by the applied field direction. A trigonal distortion of the octahedra described by the parameter $\Delta$ is expected to be finite but much smaller than the strength of the SOC in \rucl\ and  modifies the spatial magnetization density away from an ideal cubic case.}
    \label{fig:CEF}
\end{figure}

The low energy Hamiltonian of \rucl\ has been most famously described within the $j_{\textrm{eff}} = \frac{1}{2}$ pseudospin formalism. Fig. \ref{fig:CEF} shows the energy hierarchy for Ru$^{3+}$ ions in \rucl. The 4$d^5$ orbitals within an octahedral crystal field of Cl$^-$ ions split to form the $e_g$ and $t_{2g}$ orbitals. In the low-spin 4$d^5$ state (strong crystal field limit), the $e_g$ band remains unpopulated with a measured gap of $\sim$ 2.2~eV in \rucl\ \cite{koitzsch2016j,suzuki2021proximate}. The resulting Hund's coupled $t_{2g}$ band contains a single hole.
Spin-orbit coupling further splits the Hund's coupled $t_{2g}$ band. Using the hole representation, this results in a singly occupied $\Gamma_7$ doublet and unpopulated $\Gamma_8$ quartet. The ground state doublet was experimentally shown to be separated from $\Gamma_8$ by roughly 200~meV ($\sim$ 2300~K)~\cite{banerjee2017neutron,suzuki2021proximate}, meaning at our measured 80~K the system is still expected to be well-approximated by an effective $j_{\textrm{eff}} = \frac{1}{2}$ ground state. In a perfectly cubic octahedral field, the ground state wavefunction of the $j_{\textrm{eff}} = \frac{1}{2}$ state in terms of the single hole is an equal superposition of the $d_{xy}$, $d_{xz}$, and $d_{yz}$ orbitals with a complex phase describing the orbital motion~\cite{jackeli2009mott,jeong2020magnetization}. Although small, a trigonal distortion (compression) of the RuCl$_6$ octahedra has been reported~ in \rucl~ \cite{kobayashi1992moessbauer,agrestini2017electronically,lebert2020resonant,suzuki2021proximate}, shown in Fig.~\ref{fig:CEF}. This mixes the weights of the $d_{xy}$, $d_{xz}$, and $d_{yz}$ orbital constituents of what we hereafter call the generalized $j_{\textrm{eff}}=\frac{1}{2}$ state. 


To derive the magnetic form factor of \rucl\ from first principles we use a Wannier function based approach~\cite{Larson2007,walters2009effect,Xu2012,Pujol2014} that treats the trigonal crystal field and the hybridization of the magnetic cations and their surrounding anions. We perform DFT calculations within the Perdew-Burke-Ernzerhof (PBE) generalized gradient approximation (GGA)~\cite{GGA} for the exchange-correlation functional and the Linearized Augmented Plane Wave (LAPW) method as implemented in a modified version~\cite{Anton,ELKgithub} of the \textsc{elk} code~\cite{elk}, using the $R \bar{3}$ structure of \rucl\ reported in Ref. \onlinecite{mu2021role} and a $7\times7\times7$ k-mesh. We derive the Ru $t_{2g}$ Wannier functions and Wannier function based Hamiltonian of \rucl\ by projecting on the low energy bands in the [-1.5, 0.5] eV interval. While the derivation of Wannier function based Hamiltonians in the presence of SOC is implemented in our modified version of \textsc{elk}, the derivation of the spatial dependence of the Wannier functions is not. Therefore we extract the Wannier functions from a calculation without SOC and the Hamiltonian from the calculation with SOC. We note that the SOC parameter is $\lambda_{SOC}=116.82$ meV, in close agreement with experiment \cite{banerjee2016proximate,lebert2020resonant,suzuki2021proximate}, and the trigonal crystal field parameter is $\Delta=-18.47$ meV. In addition, we extract Ru $t_{2g}$ Wannier functions from an isolated Ru$^{3+}$ in a 10$^3$ \AA$^3$ box to isolate the effect of the hybridization with the Cl anions. Next, we produce the generalized $j_{\textrm{eff}} =\frac{1}{2}$ state by diagonalizing the local Hamiltonian in the presence of the Zeeman interaction with the magnetic fields corresponding to those in the neutron diffraction experiments. We then derive the magnetic form factors corresponding to the generalized $j_{\textrm{eff}} =\frac{1}{2}$ state like in Ref.~\cite{Larson2007,walters2009effect,Xu2012,Pujol2014}, where in addition to the spin magnetization density, we also compute the orbital magnetization density. We note that in general the single ion magnetization densities in \rucl\ are noncollinear~\footnote{Here by noncollinear, we mean the vector magnetization density around individual magnetic ions is not merely oriented in one direction everywhere in space}. However, in the limit that the nuclear scattering is much larger than the magnetic scattering, the magnetic structure factor (and thus form factor) extracted from Eq.~\ref{eq:flip_ratio} reduces to the projection of the transverse magnetization density along the direction of the magnetic field. Our simulations confirm that we are in this limit. Therefore, to compare against the experiment, we compute the projection of the transverse magnetization density along the direction of the magnetic field. Additional details will be presented in an upcoming manuscript \cite{TheoryFollowup}. 

\begin{figure}
    \centering
    \includegraphics[width=0.48\textwidth]{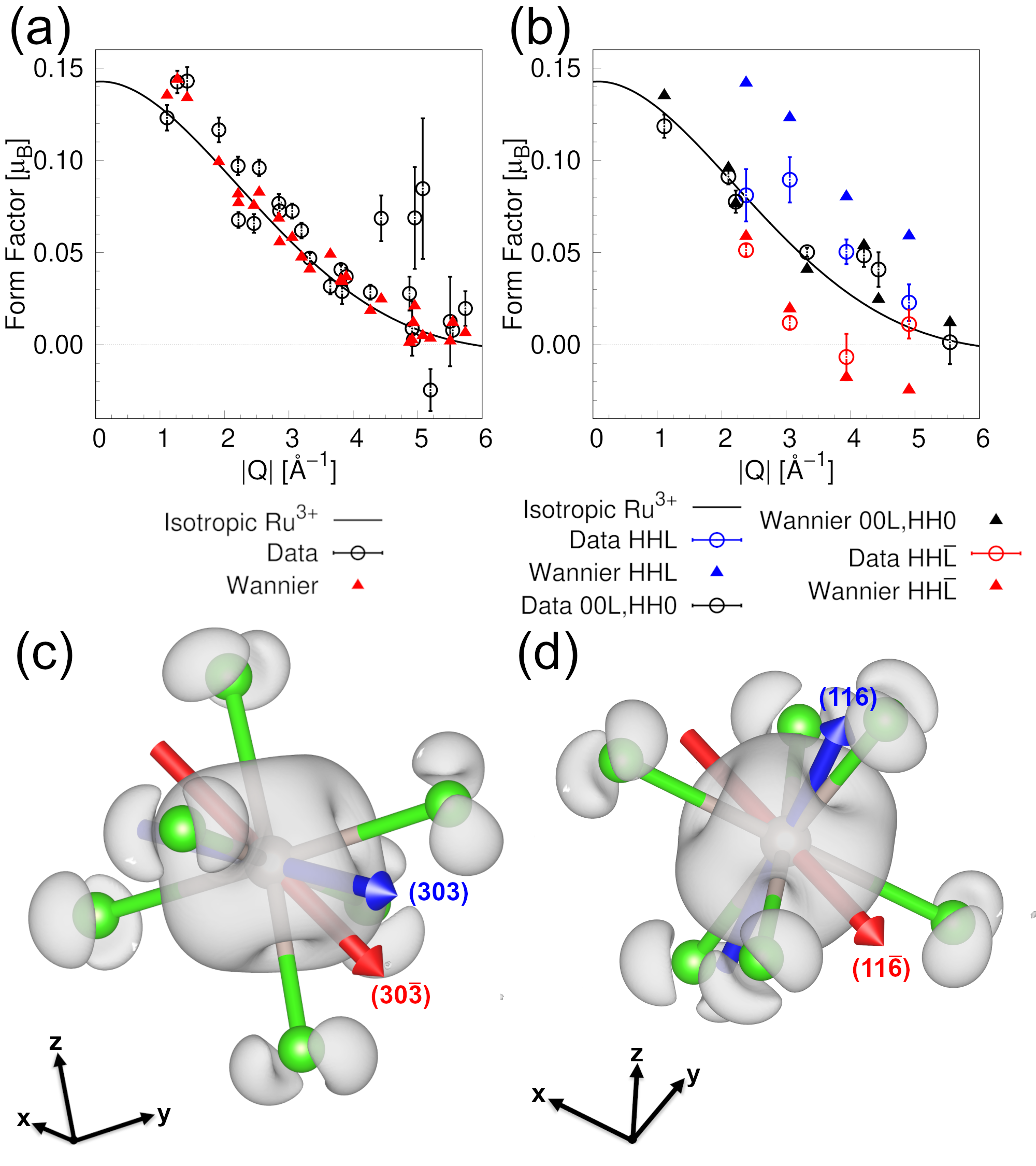}
    \caption{Wannier function based model. (a,b) Comparison between the Wannier function form factor and the ($h~0~l$) and ($h~h~l$) experiments, respectively. In (b), the anisotropic splitting between ($h~h~l$) and ($h~h~\bar{l}$) directions is apparent. (c,d) Real space spin density squared isosurfaces with relevant $\vec{Q}$ directions noted. The cartesian coordinates are the cubic axes of the local octahedron as employed in Refs.~\cite{Winter2016,eichstaedt2019deriving}.}
    \label{fig:theory}
\end{figure}

While our {\it ab initio} form factor is parameter free, we do need to introduce two parameters to compare to the neutron scattering experiment, corresponding to the ordered moment size and the domain population. By fitting these parameters against the data, we obtain a $\chi^2$ of 6.7, which is a clear improvement compared to the $\chi^2$ of 11.1 obtained from the isotropic Ru$^{3+}$ model presented in Table~\ref{tab:iso}. We also performed the fit using the same model, but with the Ru $t_{2g}$ Wannier functions replaced with those from the isolated Ru$^{3+}$ simulation. This yielded a $\chi^2$ of 7.9; which indicates that the hybridization with the Cl$^-$ anions does impact the form factor. Fig.~\ref{fig:theory}(a,b) compare form factors simulated with the Wannier function based approach against neutron scattering data taken in the ($h~0~l$) and ($h~h~l$) scattering planes respectively. The panels include the isotropic form factor obtained from the single ion Ru$^{3+}$ approximation ($\mu_{Ru}\langle j_0 \rangle_{Ru}$). In Fig.~\ref{fig:theory}(a) we see that the neutron scattering data, Wannier function based form factor, and isotropic form factor are all in reasonable agreement, with the exception of the three experimental outliers in 4-5 \AA$^{-1}$ with form factor values greater than 0.05. In Fig.~\ref{fig:theory}(b) on the other hand, we see that the experimental data and the Wannier function based form-factor display a notable splitting between the form factors along ($hhl$) and ($hh\bar{l}$) directions, a marked signature of anisotropy. The isotropic Ru$^{3+}$ model is incapable of capturing this splitting, which is the origin of the large $\chi^2$ for that model.

To better understand the nature of the splitting of the form factor branches in the ($h~h~l$) plane and lack thereof in the ($h~0~l$) plane, we present in Fig.~\ref{fig:theory}(c,d) the real-space spin density squared of the generalized $j_{\textrm{eff}} =\frac{1}{2}$ state. We note that in addition to the Ru $d$-center, the Wannier function based $j_{\textrm{eff}} =\frac{1}{2}$ spin density also incorporates Cl $p$ tails in the surrounding Cl atoms. We see that the scattering vectors (3~0~3) and (3~0~$\bar{3}$), for instance, corresponding to the point with $|Q|=3.81$~\AA$^{-1}$ in Fig.~\ref{fig:theory}(a), are along nearly symmetric directions with respect to the spin density, explaining why their form factors are essentially equal. In Fig.~\ref{fig:theory}(d) we focus on the scattering vectors (1~1~6) and (1~1~$\bar{6}$) which correspond to the widely separated points with $|Q|=3.06$~\AA$^{-1}$ in Fig.~\ref{fig:theory}(b). While (1~1~6) points along a minimum of the spin density in real space, (1~1~$\bar{6}$) points along a maximum. This explains why in momentum space the form factor along (1~1~6) is larger than along (1~1~$\bar{6}$).

\section{Conclusions}
In conclusion, we performed half-polarized neutron diffraction on single crystals of \rucl\ deep within the paramagnetic regime. By directly measuring flipping ratios of Bragg reflections within the ($h$~$0$~$l$) and ($h$~$h$~$l$) planes, we extract an experimental determination of the magnetization density and magnetic form factor of Ru$^{3+}$ in the strong crystal field limit, such as found in \rucl. Our results show an observable deviation from a simple isotropic form factor which has been been widely used in recent efforts to quantitatively describe INS data of \rucl, particularly within the ($h$~$h$~$l$) plane. While for ($h$~$0$~$l$) measurements the data appear to be reasonably well described by an isotropic form factor, we argue that including the full anisotropic form factor may provide a further constraint which may aid in determination of the complex low energy Hamiltonian in \rucl, although more investigation of the sensitivity of fit exchange parameters to inelastic neutron scattering data using full anisotropic form factor is needed. 

While we are able to describe the data on an equal footing through the single-ion multipole model or hybridized Wannier function, the differences in fits support the importance of hybridization. The main disagreement in our Wannier calculations arises from an over-representation of the ($h~h~l$) splitting, where the combination of non-equivalent domains leads to a larger systematic uncertainty in these points (see Appendix \ref{sec:Neutron}). This is in contrast to the multipole model where most of the disagreement was due to the lowest $|\vec{Q}|$ where there are only equivalently mixed (single domain) data. The large anisotropy in this region cannot be explained with the multipole and requires additional considerations which can be naturally explained through hybridization effects. Being a single ion model, our multipole model would only accurately describe the local magnetization density around the Ru$^{3+}$ ions and would likely miss features at low momentum transfers. Hybridization effects have also been recently reported in monolayer \rucl\ grown under strain on a graphite substrate\cite{wang2022direct}. This is suggestive that dealing with hybridization effects will be important not only in accurately describing the low energy Hamiltonian of \rucl, but toward implementation of device construction whether in 2D heterostructures or bulk device construction.

\begin{acknowledgements}
This research was supported by the U.S. Department of Energy, Office of Science, National Quantum Information Science Research Centers, Quantum Science Center. C.E. was supported by National Renewable Energy Laboratory, operated by Alliance for Sustainable Energy, LLC, for the U.S. Department of Energy, Office of Science, Basic Energy Sciences, Division of Materials, under Contract No. DE-AC36-08GO28308. This research used resources at the High Flux Isotope Reactor, a DOE Office of Science User Facility operated by the Oak Ridge National Laboratory. Visual representations of the crystal structure shown in this manuscript were produced using the open source software VESTA\cite{momma2011vesta}. This manuscript has been authored by UT-Battelle, LLC, under contract DE-AC05-00OR22725 with the US Department of Energy (DOE). CLS is grateful to Allen Scheie and Feng Ye for many fruitful discussions.

\end{acknowledgements}
\appendix
\counterwithin{figure}{section}
\counterwithin{table}{section}
\section{Polarized Neutron Diffraction}
\label{sec:Neutron}
Polarized neutron diffraction measurements were made at the polarized neutron triple axis instrument at the High Flux Isotope Reactor at Oak Ridge National Laboratory. For both orientations, the samples were loaded into a vertical field asymmetric cryomagnet reading a temperature of 80~K.  All measurements were taken in a three-axis mode with $E_i = E_f$ using a Heusler monochromator and PG analyzer. The Heusler alloy monochromator gave polarized neutrons with incident energies $E_i = 13.5$~meV and $E_i = 30.5$~meV. This gave a combined momentum transfer range of $|\vec{Q}|$ up to $6~\AA^{-1}$ at the elastic line. Aside from providing better optimized signal to noise, comparison of longer wavelength $E_i$ = 13.5~meV data allowed for a limit on energy dependent corrections to intensities such as from extinction corrections described below. Two experiments were taken on two separate high-quality single crystals of \rucl\ which measured reflections in the ($h$~$0$~$l$) and ($h$~$h$~$l$) scattering planes.

Polarized neutrons scatter through interactions with both nuclei and local magnetization. For simple co-linear magnetic system the total scattering cross-section of the neutron is given by
\begin{equation}
    \sigma = F_{N}F_{N}^* + \vec{F}_M^{\perp} \cdot \vec{F}_M^{\perp *} + \vec{P} \cdot (F_{N}\vec{F}_M^{\perp *} + \vec{F}_M^{\perp} F_{N}^* )
    \label{eq:cross}
\end{equation}
where $F_N$ is the nuclear structure factor and $\vec{F_M^{\perp}}$ is known as the magnetic interaction vector, related to the magnetic structure factor by $\vec{F}_M^{\perp} = \hat{Q} \times \vec{F}_M \times \hat{Q}$.  Using a half-polarized setup, the incident neutron is polarized to be along the direction of an applied magnetic field. The neutron spin state is then controlled by a polarization flipper to either be directed parallel or antiparallel to the applied field. The resulting scattered intensities are not sensitive to polarization and contain both nuclear and magnetic contributions. As Eq. \ref{eq:cross} shows, only the component of the magnetization density perpendicular to the scattering plane contributes to the scattering. Therefore we are only sensitive to a projection of the magnetic structure factor and thus a projection of the overall magnetization density. 

To get from Eq. \ref{eq:cross} to Eq. \ref{eq:flip_ratio} in the main article, we make two assumptions: (1) the measured magnetization density is a scalar magnetization density which is oriented orthogonal to the scattering plane and (2) the nuclear (and thus $\vec{q}$ = 0 paramagnetic) structure factor is real valued, as required by centrosymmetry. While our experimental setup and measurements deep into the paramagnetic regime may justify the first assumption, the second assumption is less certain for \rucl. As discussed in the main text, the exact nuclear unitcell is not uniquely determined due to domains and sample dependency. As of the time of this publication, much debate is ongoing between samples grown via vapor transport and modified Bridgeman growth which seem to show strong inconsistencies in their low temperature properties. We note that this may contribute a larger systematic uncertainty beyond what is reported in this experiment, where we fixed the nuclear unitcell as a given parameter. Nevertheless, the best-fit low temperature $R~\bar{3}$ structure used in this work is consistent with a centrosymmetric unitcell and (2) can be justified. With these assumptions $M_\perp \xrightarrow[]{}\ pF_M$ when the flipper is not used ($I_{+}$) and $M_{\perp} \xrightarrow[]{}\ \epsilon p F_M$ with the flipper ($I_-$). Here $p$ and $\epsilon$ represent the efficiencies of the polarizer and flipper respectively.   

The flipping ratio technique allows for a systematic elimination of many errors associated with complexities such as absorption effects and Debye-Waller factors (for measurements taken at the same temperatures). Additionally when compared to a full polarization analysis, the half-polarized determination of the flipping ratio helps to minimize errors in the extracted magnetic structure factor arising from the efficiency of the polarizer ($p$). Rather than being sensitive to the flipping ratio of the polarizer ($\frac{1-p}{1+p}$), the magnetic structure factor is only sensitive linearly to $p$. Despite these features, an accurate determination of the magnetic structure factor requires a thorough accounting of sources of instrumental uncertainties. With this in mind, we estimated sources of uncertainty in our experiment due to the initial beam polarization efficiency ($p$), flipper efficiency ($\epsilon$), extinction, $\lambda/2$ contribution, and multiple scattering effects.

To estimate the efficiency of both the initial beam polarization and flipper, we installed a Heusler alloy analyzer on the scattered side and set it to measure the non-spin-flip channel. We then took measurements of the flipping ratios of a non-magnetic aluminum powder ring coming from our sample mount. These measurements were taken at both incident energies, $E_i = 13.5$~meV and $E_i = 30.5$~meV for various field strengths, in order to maximize the observed flipping ratio. Vertical guide fields are field dependent and were chosen at measured values which maximized measured flipping ratios at the aluminum Bragg peak. Since the strength of the magnetic scattering from our sample will be determined by the size of the field polarized moment (and hence magnetic field strength), a maximal field of 8~T and 6~T were chosen after confirming the field did not have a significant effect on the beam polarization. This resulted in vertical guide fields of 3.0~T and 3.2~T for all flipper measurements in the ($h$~$0$~$l$) plane (8~T) and ($h$~$h$~$l$) plane (6~T), respectively. Aluminum peaks gave the combined product of the initial beam polarization efficiency and the flipper efficiency. The flipper efficiency was separately estimated be 0.99(6) for both energies and insensitive to the field strength, leaving the beam polarization efficiencies for our experiments as follows:
\begin{equation}
\begin{split}
      p = 0.84 \pm 0.03~~(E_i = 13.5~meV, \vec{B} = 8~T)\\ 
      p = 0.81 \pm 0.06~~(E_i = 30.5~meV, \vec{B} = 8~T)\\
      \\
      p = 0.84 \pm 0.04~~(E_i = 13.5~meV, \vec{B} = 6~T)\\ 
      p = 0.82 \pm 0.05~~(E_i = 30.5~meV, \vec{B} = 6~T)\\
\end{split}
\end{equation}

To eliminate $\lambda/2$ contributions to the scattering, PG filters were installed for all measurements. Despite this, some contribution is expected. An estimate of the size of this contribution can be made by looking at a forbidden peak which corresponds to one of the brightest reflections for $\lambda/2$). By comparing the intensities of ($1$~$1$~$3$) with ($0.5$~$0.5$~$1.5$) we constrained contributions from $\lambda/2$ to be $<$ 1\%. A limit on the presence of multiple scattering in the ($h$~$0$~$l$) plane geometry was estimated by looking at the combination of two relatively strong reflections $(1~0~1)$ and $(3~0~0)$ as well as the weak $(4~0~1)$ reflection. By comparing the calculated vs measured intensities in zero-field, we can look for an increase in the weak $(4~0~1)$ intensity in order to quantify the contribution of multiple scattering. We also checked the weak reflection $(1~0~13)$ compared to the relatively bright $(0~0~12)$ and $(1~0~1)$ reflections. We find no evidence of a deviation in the intensities beyond measured experimental uncertainties ($\leq$  2\%) for either Bragg peak and conclude contributions from multiple scattering in our experiment were not likely to cause large systematic uncertainties in the extracted magnetic structure factor for the ($h$~$0$~$l$) reflections. For the ($h$~$h$~$l$) geometry, such a contribution was unable to be uniquely estimated due to the mixture of structural domains. However we note the between the combination of multiple very bright Bragg reflections and the high symmetry leading to many combinations of reflections within this plane, it is possible multiple scattering is non-negligible for ($h$~$h$~$l$) reflections and contribute to a larger systematic error in our measurements beyond our reported uncertainty. 

Since we investigated relatively large single crystals with longer wavelength neutrons, a careful treatment of extinction was necessary. We treated extinction according to a first order expansion of spherical domains with a Gaussian angular mosaic distribution following the formalism of Becker and Coppens applied to polarized neutron diffraction~\cite{becker1974extinction,schweizer1980polarised}. Under this approximation the form of the flipping ratio is modified slightly to

\begin{equation}
    R = \frac{F_{N}^2 + 2psin^2(\alpha)F_{N}F_{M} + sin^2(\alpha)F_{M}^2 + ES^+}{F_{N}^2 - 2p\epsilon sin^2(\alpha)F_{N}F_{M} + sin^2(\alpha)F_{M}^2 + ES^-}
    \label{eq:extinction}
\end{equation}
where the extinction terms $ES^+$ and $ES^-$ are given by 

\begin{equation}
\begin{split}
    ES^+ = -\frac{1}{2} \left[ (f^{+2})^2(1+p)ES + (f^{-2})^2(1-p)ES \right]\\
    ES^- = -\frac{1}{2} \left[ (f^{+2})^2(1-p\epsilon)ES + (f^{-2})^2(1+p\epsilon)ES \right]
\end{split}
\end{equation}
with
\begin{equation}
    ES = \frac{\lambda^3}{v_0^2 sin(2\theta)} \left(\frac{2\tau^2 sin(2\theta)}{3\lambda} + \frac{\bar{T}}{\sqrt{[\lambda/\tau sin(2\theta)]^2 +1/2g^2}}    \right)
\end{equation}
\begin{equation}
\begin{split}
    f^{+2} = F_N^2 + sin^2(\alpha)F_M^2 + 2sin(\alpha)F_NF_M\\
    f^{-2} = F_N^2 + sin^2(\alpha)F_M^2 - 2sin(\alpha)F_NF_M 
\end{split}
\end{equation}
Here $\tau$ corresponds to the average size of perfect domains, $\bar{T}$ describes the average flight path of a neutron through the sample, $\lambda$ is the wavelength of the neutron, and  $g$ represents the intrinsic mosaic distribution of perfect domains. A careful inspection of Eq. \ref{eq:extinction} shows the quadratic dependence of $R$ on $F_{m}$ becomes quartic with the addition of extinction, but for the case where $F_M \ll F_N$ one can approximate Eq. \ref{eq:extinction} well by converting to $y = F_M/F_N$ and keeping only up to linear terms in $y$. For our experiment, we remained heavily in the $F_M \ll F_N$ regime for all measured values, with differences in extracted values of $F_M$ differing less than 2\% for all measured values between linear and full quartic treatments. 

While this extinction model may not accurately describe the type of extinction found in the 2D van der Waals \rucl, it can provide an estimate of the effect of extinction. For secondary extinction, values of $g$ were able to be estimated via direct Bragg peak widths, as both crystals featured mosaics well beyond instrument resolution (FWHM $\sim$ 5$^\circ$ for sample 1 and FWHM $\sim$ 2$^\circ$ for sample 2). Both crystals were hemispherical shaped and $\Bar{T}$, the average neutron flight path through the sample was estimated as half that of a spherical sample ($\bar{T}_{sphere}$ = 3R/2). Values of $\tau$, the average size of perfect spherical grains were less well constrained. Since the stacking faults induce a separate magnetic order, a lower limit of $\tau$ was estimated through relative brightness of the two magnetic order parameters. Inclusion of stacking faults, which along with the primary magnetic transition at $T \sim 7$~K into a $\vec{q} = (1/2~0~1)$ ordered phase, results in a second magnetic transition at $T \sim 14$~K into a $\vec{q} = (1/2~0~3/2)$ ordered phase. For both samples used in this experiment, we tracked the zero-field order parameter corresponding to these two magnetic orders. We no find intensity past background at the $(1/2~0~3/2)$ Bragg peaks above $\sim 1\%$ the intensity observed at the corresponding $(1/2~0~1)$ peak. From this we conclude that our samples do not have a large volume fraction of stacking faults, which we used as a lower bound of $\tau$ as 100 times the crystallographic $c$-axis (1700 $\AA$). While we find extinction corrections to be non-negligible, the difference in extinction for each neutron spin state is small for most measured flipping ratio and thus corrections to our flipping ratio values remain small. This was particularly true for sample ``A'', where the measured reflections show no dependence on extinction within experimental error. 

\begin{figure}
    \centering
    \includegraphics[width=\columnwidth]{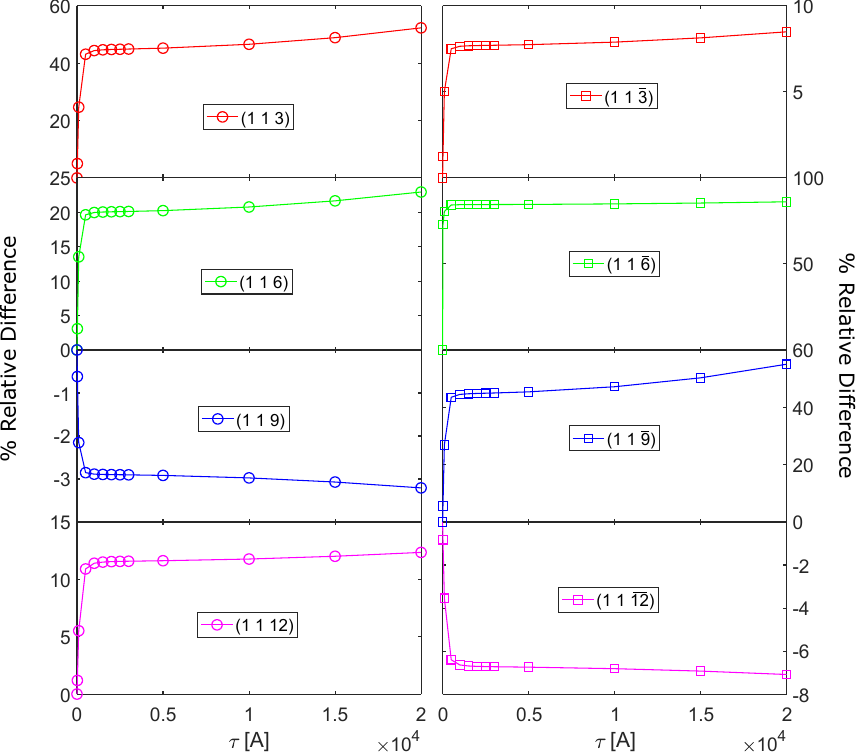}
    \caption{Dependence of ($h$~$h$~$l$) data to $\tau$, the average size of perfect domains in extinction corrections. Due to mixture of two nonequivalent reflections with different relative extinction corrections, data is sensitive to small extinction corrections. For reasonable ranges around $\tau \sim 10^{3}~\AA$, the values remain fairly stable. }
    \label{fig:tau_dep}
\end{figure}

The mixed reflections which combine ($h$~$h$~$l$) and  ($h$~$h$~$\bar{l}$)  with both $h$ and $l$ nonzero can vary strongly with inclusion of extinction depending on the relative domain populations. The dependence of these reflections to the unknown extinction parameter $\tau$ are shown in Fig. \ref{fig:tau_dep} for a domain population of 95/5 and a fixed secondary extinction parameter $g$ of 24.4. While we observe a strong deviation in data for any non-zero estimate of extinction, reasonable ranges of $\tau \leq 20000~\AA$, do not affect the data strongly. This indicates that extinction for crystal ``B'' is mostly dominated by secondary extinction, where the relevant parameter $g$ can be safely estimated.

Surprisingly, the data in \emph{both} experiments show a negative induced magnetic moment consistent for \emph{all} measured values of $F_M$. While our experimental setup has ruled out a simple explanation for this observation, we nevertheless attribute this to a consequence of our experimental definitions given the large collective data showing paramagnetic rather than diamagnetic behavior in \rucl. This experimental detail does not affect our analysis aside from an overall negative sign, as one can simply invert the direction of the field (and thus the observed flipping ratios) to recover a positive moment. All data presented in the main article show the inverted measured flipping ratios such that the resulting induced moment is paramagnetic.

\subsection{Second Structural Domain}
As mentioned above, during our ($h~0~l$) experiment below the $1^{st}$ order structural phase transition, we observed scattering consistent with a second structural domain. This second domain followed the reflection conditions: $h$ + $l$ = 3$n$ ($k$ = 0) in the hexagonal $R \bar{3}$ setting. For many reflections in the ($h$~$0$~$l$) plane, most measured reflections correspond to forbidden peaks in one of the two domains, however reflections of the type ($3n$~$0$~$3n'$) where $n, n'$ are integers, are allowed by both. This means that measured flipping ratios belonging to these reflections contain contributions from both domains. While the nuclear structure factor is identical for both domains, the vector nature of the magnetic structure factor is less clear. In this case, the global applied magnetic field corresponds to a different (albeit symmetry related) direction for each domain, which may induce a different magnetization due to anisotropy in the material. 

By measuring flipping ratios and intensities of multiple reflections where only one domain was present, we were able to check for any evidence of mixing effects in the reflections allowed for both domains. Table \ref{tab:domain} shows comparisons between the two domains for multiple reflections. We observe similar scattered intensities which indicate similar populations of each domain. We also see no deviation of measured flipping ratios within our experimental uncertainty. Since the nuclear contribution is identical by choice of unitcell, this implies that the magnetic contributions are also identical. This allowed us to separate the contributions from each and accurately model our data assuming a single domain approach.

\begin{figure}
    \centering
    \includegraphics[width=0.85\columnwidth]{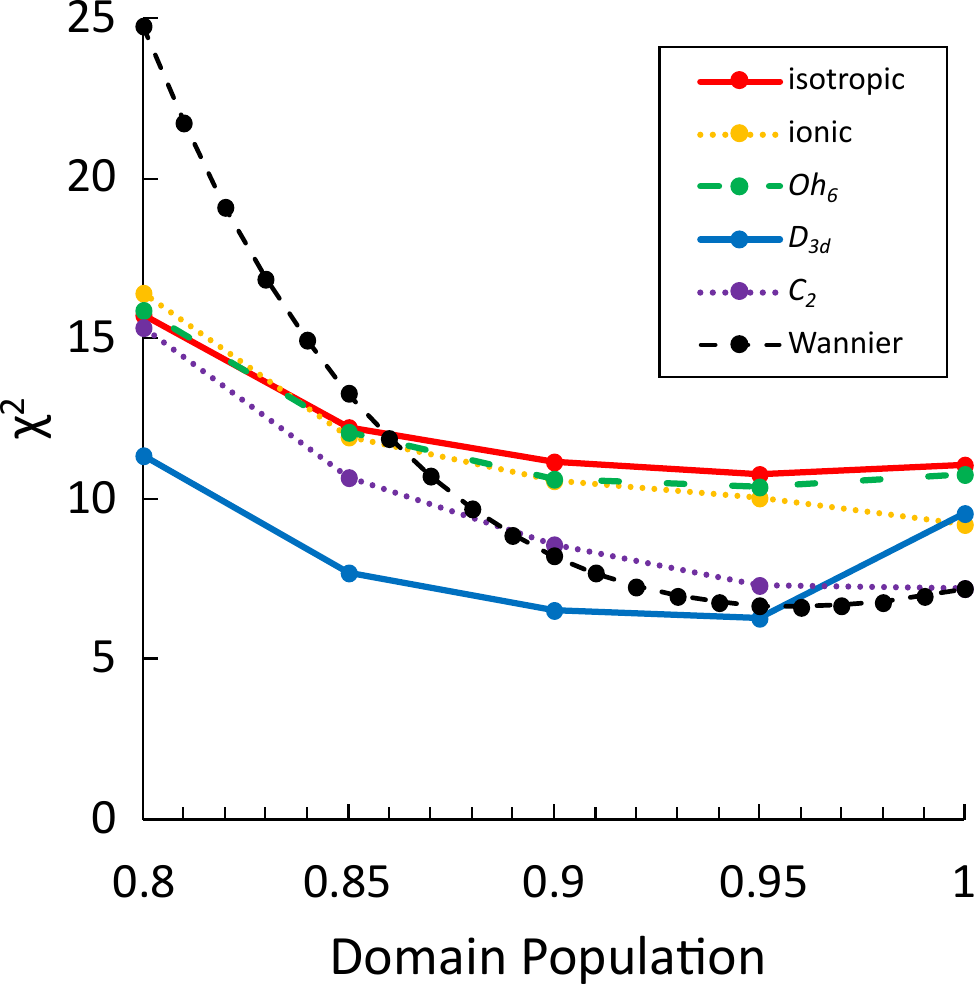}
    \caption{Relative best-fit $\chi^2$ as a function of domain populations for different models. For each model considered in this work, the lowest best-fit $\chi^2$ is found in the vicinity of 0.95 $\pm$ 0.05, indicating that sample ``B'' was nearly monodomain. }
    \label{fig:domain_pop_fits}
\end{figure}

For measurements in the ($h$~$h$~$l$) plane, we were unable to directly observe a secondary structural domain. Because all measured reflections are allowed by both reflections, each corresponding flipping ratio contains a contribution from both domains. As in the case for ($h$~$0$~$l$) measurements, reflections with either $h$ = 0 or $l$ = 0 mix equivalent reflections and were used to constrain extinction effects described above. Although we originally focused first on a nearly equal domain population case,  when we used this domain population estimate we found resulting values $F_M$ were \emph{unphysical} (with close to 50/50 domain populations give $\mu \cdot F_M^{119} ~ 5\mu_B $). In fact the only way to push $F_M$ toward reasonable values was to further increase the domain population close to the monodomain situation. We settled on the final domain population of  0.95(5)/0.05(5) after finding that \emph{all} best fits to the data as a function of domain population show minima in the range of 95\%. A plot of the relative $\chi^2$ best fit values as a function of domain population is shown in Fig. \ref{fig:domain_pop_fits}. For each model a shallow minima in the vicinity of 0.95 is observed. While the flipping ratios eliminate absorption and Debye-Waller factors, the relative domain populations inferred from 0T integrated Bragg peak intensities mentioned above do not. This may contribute toward the extreme disagreement in estimated domain populations between the pair of nuclear Bragg peak intensities and that found by comparing the magnetic structure factor to models. The large deviation away from equal populations may be due to the lack of past thermal cycling in sample ``B'' combined with a careful slow cooling protocol we used while crossing the structural transition during the experiment. From $T$ = 175~K to $T$ = 110~K we lowered the temperature at roughly 1~K/min, whereas for sample ``A'' the temperature was lowered without careful treatment of the cooling rate.

\begin{table}[]
\caption{Table showing extracted normalized intensities and flipping ratios for several measured reflections belonging to the primary ($p$) and secondary ($s$) domains. Within experimental uncertainty we see good agreement in intensities and identical flipping ratios.}
\label{tab:domain}
\begin{tabular}{ccccc}
\hline
                & \begin{tabular}[c]{@{}c@{}}Domain\\ p:($-h + l = 3n$)\\ s:($h + l = 3n$)\end{tabular} & \begin{tabular}[c]{@{}c@{}}$I_{-}$\\\end{tabular} & \begin{tabular}[c]{@{}c@{}}$I_{+}$\\ \end{tabular} & $R$ \\ \hline
$(1~0~1)$       & p                                                                                             & 0.2237(7)                                                      & 0.2566(7)                                                      & 1.147(4)       \\
$(1~0~\bar{1})$      & s                                                                                             & 0.2502(7)                                                      & 0.2869(7)                                                      & 1.147(3)       \\ \hline
$(\bar{2}~0~4)$ & p                                                                                             & 0.1269(5)                                                      & 0.1370(5)                                                      & 1.080(5)       \\
$(2~0~4)$       & s                                                                                             & 0.1025(4)                                                      & 0.1098(5)                                                      & 1.072(6)       \\ \hline
$(\bar{1}~0~2)$ & p                                                                                             & 0.2106(6)                                                      & 0.2408(7)                                                      & 1.161(4)       \\
$(1~0~2)$       & s                                                                                             & 0.1512(8)                                                      & 0.1767(8)                                                      & 1.168(6)       \\ \hline
$(4~0~1)$       & p                                                                                             & 0.0534(3)                                                      & 0.0547(3)                                                      & 1.02(1)        \\
$(\bar{4}~0~1)$ & s                                                                                             & 0.0535(3)                                                      & 0.0539(3)                                                      & 1.00(1)       
\end{tabular}
\end{table}

\subsection{Sample Dependence and Combining Different Orientations}
For \rucl, a long standing complication has been sample dependence of material properties. This has been has been observed in many aspects of \rucl\ properties, but the extent of sample dependence of neutron scattering data has not been thoroughly investigated. The possibility of sample dependency means a careful analysis is required in order to combine data between two separate samples. The primary cause of sample dependence has been related to stacking faults. As mentioned above, for both samples used in this experiment, we tracked the zero-field order parameter corresponding to these two magnetic order in order to quantify that our samples do not have a large volume fraction of stacking faults.

In order to further look at any sample dependence, we look at the observed flipping ratios and the resulting form factors of the  ($0$~$0$~$l$) reflections, which are observable in both orientations, and thus both samples. First we take into account the difference of magnetic field orientations with respect to the two orientations. While there are reports of a strong anisotropic $g$-tensor for \rucl\cite{agrestini2017electronically,kubota2015successive}, all groups only report an anisotropy between in-plane and out-of-plane components, $g_{ab}$ and $g_c$ expected for $C^{*}_{3}$ symmetry\cite{balz2021field}. While there is in-plane anisotropy present due the combination of deviations away from $C^{*}_{3}$ symmetry and anisotropic exchange, small differences in magnetic fields applied within the honeycomb plane necessary to disrupt magnetic order (7.3~T vs 8~T at 2~K) suggest the anisotropy is small. In particular, since both measurements were taken at 80~K, deep into the paramagnetic phase, we do not expect a stark difference in flipping ratios between the two orientations.

\begin{figure}
    \centering
    \includegraphics[width=\columnwidth]{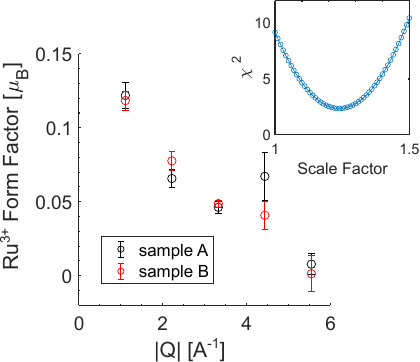}
    \caption{(Left) Extracted raw form factors (not normalized to be 1 such that they include the induced moment) of shared  ($0$~$0$~$l$) reflections for both experiments, where sample ``B'' has been scaled to match the difference in applied fields between the two experiments. Inset shows $\chi^2$ of the scaling factor between the two sets of data, showing a clear minimum around a scale factor value of 1.24. The data shows good agreement between two samples. We attribute the disagreement observed in the $(0~0~12)$ reflection ($|\vec{Q}| \sim 4.4~\AA^{-1}$) to low counting statistics measured for this reflection in the  ($h$~$0$~$l$) experiment.}
    \label{fig:comparison}
\end{figure}

Figure \ref{fig:comparison} shows the extracted raw form factors of the ($0$~$0$~$l$) reflections between the two experiments, assuming a local moment on Ru$^{3+}$ ions. Since we applied a maximum field of 8~T for all ($h$~$0$~$l$) measurements and a 6~T field for ($h$~$h$~$l$) measurements, we expect a different induced moment and can scale observed $F_{m}$ between shared ($0$~$0$~$l$) reflections to provide a direct scaling between measurements made across the two experiments. We find a best fit scaling factor of $1.24 \pm 0.06 $ which is in good agreement for a paramagnetic induced moment in a 6T vs 8T field. When scaled, the out of plane reflections appear consistent within measurement error. The disagreement in the reflection ($0$~$0$~$12$) ($|\vec{Q}| \sim 4.4~\AA^{-1}$), we attribute to the low statistics measured in the ($h~0~l)$ data reflecting its relatively large statistical uncertainty.

\section{Comparison of Model Fits}
All models were fit using a least squares/maximum likelihood approach using a generalized reduced $\chi^2$ function of the form
\begin{equation}
    \chi^2 = \frac{1}{N-P}\sum_{i=1}^{N} \frac{(F_{i}^{calc}-F_{i})^2}{\sigma_{i}^2}
\end{equation}
where $N$ correspond to the set of independent extracted magnetic structure factors for each measured point in reciprocal space and $P$ is the number of parameters used in fitting, such that the function is normalized by the relative degrees of freedom. This gave us a way to compare fits of models with differing numbers of parameters.

During our $(h$~$h$~$l)$ experiment, due to experimental limitations, flipping ratios of four reflections were measured corresponding to mixtures of primary and twinned reflections whose opposite flipping ratio could not be measured (for example, $(2$~$2$~$3)$ and $(2$~$2$~$\bar{3})$. For most models since we were unable to independently extract the primary and twinned structure factors, we did not include this data in the analysis. Calculation of magnetization using Wannier functions however, were done \emph{ab~initio} and thus did not require the knowledge of each $F_M$. Instead calculation of the flipping ratios could be done directly using known domain populations in order to compare with these points. The measured flipping ratios of these reflections are included here for completeness (see Table \ref{tab:data_hhl_2}).

\begin{table}[]
\caption{Experimental data of flipping ratios for the $(h$~$h$~$l)$ plane in \rucl. Due to lack of information corresponding to the flipping ratio of the twinned reflection, extraction of $F_M$ was not possible for these data. Instead comparison of the observed flipping ratios directly can be used to further constrain fits to Wannier functions. All data is taken at 80~K and with a 6~T applied field along a $\{1$~$0$~$0\}$ equivalent direction.}
\label{tab:data_hhl_2}
\begin{tabular}{cccc}

\hline
$(h~k~l)$~~~        & $R$~~~         & $F_{N}$ [fm]~~~        \\ 
\hline

$(2~2~\bar{3})$~~~  & 1.00(2)~~~     & -178.789~~~            \\
$(2~2~\bar{6})$~~~  & 0.99(2)~~~     & 151.561~~~             \\
$(2~2~\bar{9})$~~~  & 1.01(2)~~~     & 73.705~~~              \\
$(1~1~\bar{15})$~~~ & 1.011(5)~~~    & 128.762~~~             \\\hline
\end{tabular}

\end{table}

\section{Multipole Expansion}
\label{sec:multipole}

\begin{figure}
    \centering
    \includegraphics[width=\columnwidth]{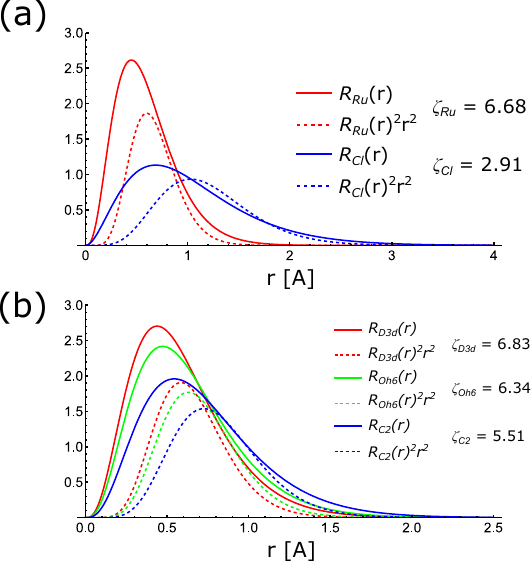}
    \caption{ Direct-space radial distributions for simple ionic isotropic limit best-fit radial integrals presented in the main article. (b) Direct-space radial distributions for best-fit multipole fits. While the value of $\zeta$ varies slightly, all models show that the radial extent of the magnetization densities are constrained to be less than $\sim$ 2 $\AA$. }
    \label{fig:radial}
\end{figure}

\begin{figure*}
    \centering
    \includegraphics[width=\textwidth]{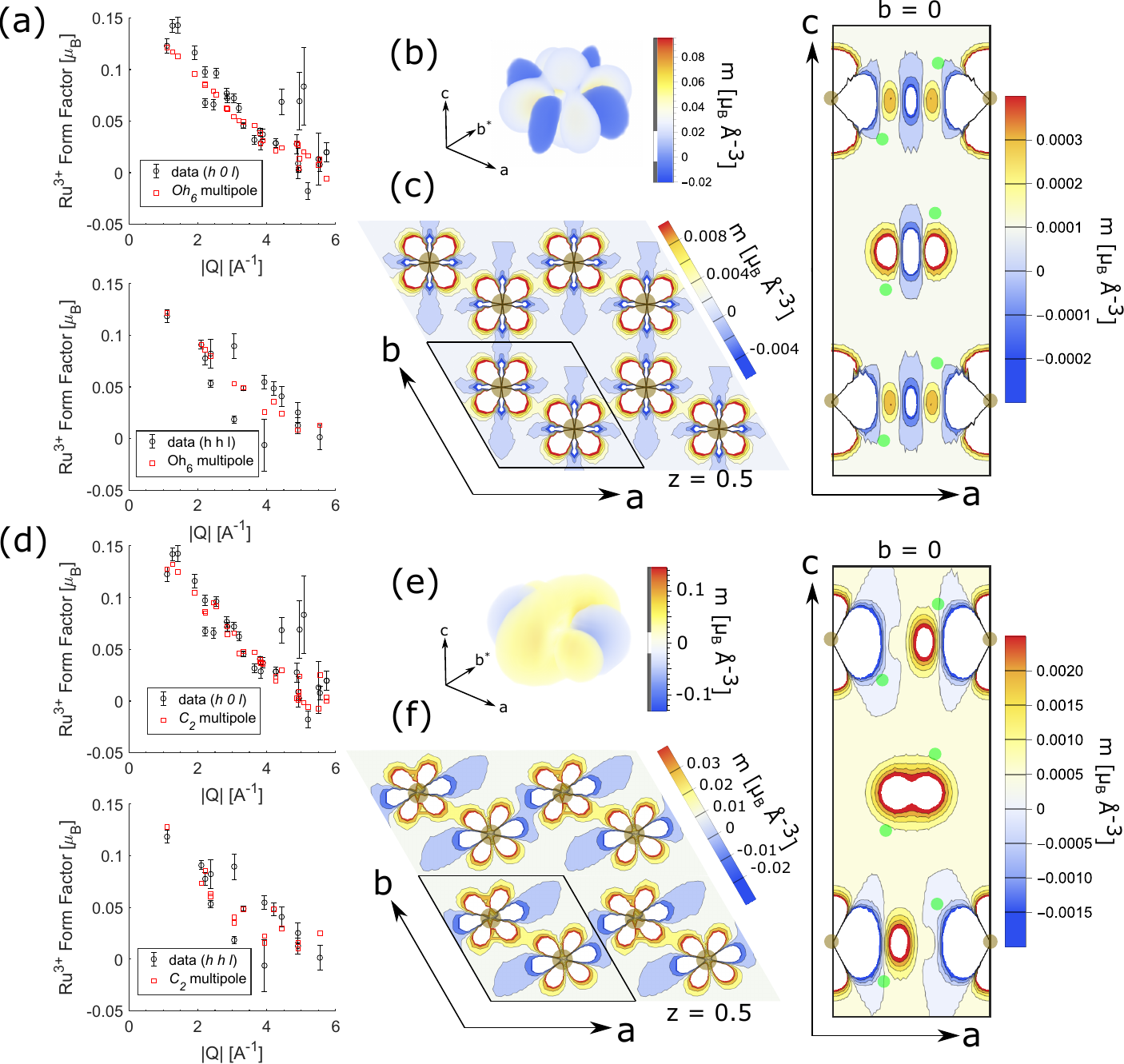}
    \caption{(a) Best fit $Oh_{6}$ multipole model compared to the unnormalized single ion Ru$^{3+}$ form factor for data taken in the ($h~0~l$) plane and ($h~h~l$) plane. (b) Direct space scalar magnetization density for best fit $Oh_{6}$ multipole model magnetization density centered described in (a) on a single Ru ion. (c) Contour plots of the the $Oh_6$ magnetization density shown above in the $ab$ plane centered on an optimal Ru honeycomb layer and $ac$ plane showing out of plane extent of the magnetization density. The brown and green spheres show Ru and Cl sites in the unitcell, respectively. (d) $C_{2}$ multipole model compared to the unnormalized single ion Ru$^{3+}$ form factor for data taken in the ($h~0~l$) plane and ($h~h~l$) plane. (e) Direct space scalar magnetization density of $C_{2}$ multipole model magnetization density described in (d) for a single Ru ion. (f) Contour plots of the the $C_2$ magnetization density shown above in the $ab$ plane centered on an optimal Ru honeycomb layer and $ac$ plane showing out of plane extent of the magnetization density }
    \label{fig:multipole_supplemental}
\end{figure*}

In principal, any arbitrary magnetization density can be constructed using a multipole expansion~\cite{coppens1997x,boothroyd2020principles}. This provides a useful model one can parameterize to describe complex magnetization densities, such as systems with strong spin and orbital magnetization. For a scalar magnetization due to a single magnetic ion, the expansion can be expressed as
\begin{equation}
    m(\vec{r}) = \sum_{l}R_l(r) \left[a_{l0}Z_{l0}(\hat{r}) + \sum_{m = l}^l \left( a_{lm}^{c}Z_{lm}^{c}(\hat{r}) + a_{lm}^{c}Z_{lm}^{c}(\hat{r})\right) \right]
    \label{eq:multipole_m}
\end{equation}
where $a_{lm}$ are fitting parameters, $R_{l}(r)$ are functions describing the radial distribution of the magnetization density, and the angular components are described in terms of Tesseral Harmonic functions, $Z_{lm}$. These are real combinations of spherical harmonic functions $Y_{l,m}$ defined as
\begin{subequations}
\begin{align}
    Z_{l0} &= Y_{l,0}\\
    Z_{lm}^{c} &= \frac{1}{\sqrt{2}} \left( Y_{l,-m} + (-1)^{m}Y_{l,m} \right)\\
    Z_{lm}^{s} &= \frac{i}{\sqrt{2}} \left( Y_{l,-m} - (-1)^{m}Y_{l,m} \right)
\end{align}
\end{subequations}
Note that for Tesseral Harmonic functions $m$ $>$ 0. Expansion of exp$(i \vec{Q} \cdot \vec{r})$ allows a simple Fourier transform of Eq. \ref{eq:multipole_m}. The resulting magnetic structure factor of a single magnetic ion is then given by
\begin{equation}
    F_{M}(\vec{G}) = 4\pi p_0 \sum_{l}i^{l}\braket{j_{l}}\left[\Tilde{a}_{l0}Z_{l0} + \sum_{m = 1}^l \left( \Tilde{a}_{lm}^{c}Z_{lm}^{c} + \Tilde{a}_{lm}^{s}Z_{lm}^{s}\right) \right]
    \label{eq:multipole_fm}
\end{equation}
here $\braket{j_{l}(G)}$ are radial integrals of the radial component of the magnetization and spherical Bessel functions, $j_l$, given by $\braket{j_{l}} = \int R(r)^2j_{l}(Gr)r^2dr$. $\vec{G}$ correspond to the vector in reciprocal space such that $\hat{G}$ describes the spherical angles of the Tesseral Harmonic functions. For Ru$^{3+}$ in \rucl\ these directions either naturally correspond to vectors perpendicular to Ru-Cl-Ru bonding planes (no trigonal distortion) or $\hat{s_z}$ along the direction of the trigonal distortion (crystallographic $c$-axis). In latter case, we define $\hat{s_x}$ to be along the crystallographic $a$-axis such that $\hat{s_x}$ points perpendicular to a Ru-Ru bond. We note here that we also tested our fits with $\hat{s_x}$ along the crystallographic $a^*$ axis (parallel to a Ru-Ru bond), however this definition always lead to a worse relative fit to the data. Extending to multiple magnetic ions per unit cell, the magnetic structure factor simply becomes the sum of the single ion magnetic structure factors taking into account phase differences.
\begin{equation}
    F_{M}(\vec{G}) = \sum_{j} \left[ F_{M,j}e^{i \vec{G} \cdot \vec{r}_j} \right]
\end{equation}
For our fits we assume a multipole expansion centered on only Ru$^{3+}$ ions, i.e. we ignore hybridization with Cl$^{-}$ ions. 

The radial function was defined following Ref. \onlinecite{jeong2020magnetization}, as a simple normalized Slater-type function of the form
\begin{equation}
    R_n(r) = \sqrt{\frac{(2 \zeta)^{2n+1}}{(2n)!}}r^{n-1}e^{-\zeta r}
\end{equation}

with $n$ = 4 for the 4$d^5$ orbitals of Ru$^{3+}$ and $\zeta$ being a fit parameter describing the radial distribution. To test whether the radial distribution for the spin and orbital components differed, we tested fits using two separate $\zeta$ parameters against fits using a single $\zeta$ parameter. We compared fits to both our data and previously used isotropic Dirac-Slater calculations~\cite{cromer1964alamos} focusing on the low $|\vec{Q}|$ data. In the case of the Dirac-Slater calculations we model the Ru$^{3+}$ using the unfilled $d$-orbital contributions found in Ref. \onlinecite{cromer1964alamos}.

When comparing to the Dirac-Slater calculations, we obtain a marginally better fit to two radial components comparing reduced-$\chi^2$ statistics. With two parameters a broad minimum in $\chi^2$ is always with respect to one parameter, indicating that the fit is mostly sensitive to only one of the radial parameters. When trying to fit our experimental normalized form factor derived using Eq. \ref{eq:form_factor} in the main text, the larger spread in values dominated the $\chi^2$ statistic. In this case, a single radial parameter was able to produce a lower reduced-$\chi^2$ than two radial parameters. Along with greatly simplified calculations, this lead us to restrict our analysis to a single radial parameter. We find relatively good agreement in radial parameters across fits both to Hartree-Dirac-Slater calculations and data using all models. We find best fit parameters of $\zeta$ range between  $\zeta \sim 5.5~\AA^{-1}$ and $\zeta \sim 6.8~\AA^{-1}$. The direct-space radial distributions are shown in Fig. \ref{fig:radial}.

While the sum in Eq. \ref{eq:multipole_fm} is infinite in principle, the local symmetry of the magnetic ions dictate which coefficients are non-zero. Looking at radial integrals of our best-fit radial function (presented in Fig. \ref{fig:isotropic}b of the main text), it is clear that for experimentally achieved range of $|\vec{Q}|$, we are only sensitive up to $l$ = 4 terms of the multipole expansion. Constraining ourselves strictly to comparisons with the $d$-orbitals limits the series to $l$ = even terms, due to the local inversion symmetry of $d$-orbitals. For \rucl\ the large splitting of the $t_{2g}$ and $e_g$ bands should limit the magnetization density to be dominantly described by combinations of the $t_{2g}$ band orbitals (in the case of no $p$ hybridization with Cl). In this work we considered four site symmetries: $Oh_6$, $D_{3d}$, $C_3$, and $C_2$. For comparison to overall best-fit $D_{3d}$ model in the main article, we also show the best-fit models assuming $Oh_6$ and $C_2$ symmetry in Fig. \ref{fig:multipole_supplemental}.

In the case of a perfect cubic octahedral environment, the only non-zero terms up to $l$= 4 are ($Z_{00}$,  $Z_{40}$, and $Z_{44}^c$) due to the high symmetry. As a confirmation that the observed magnetic structure factor could not be reproduced solely within a $Oh_6$ single ion multipole representation, we also included two additional symmetry allowed terms corresponding to $l$ = 6, namely $Z_{60}$ and $Z_{64}^c$. Up to $l$ = 6 we found no better fit to our data resulting in a worse overall fit due to the reduced degrees of freedom. The resulting form factor is close in form to the isotropic single ion $\braket{j_0}$ described in the main article, fitting mostly the center of mass of the data. Slight deviations in the high $|\vec{Q}|$ due to $Z_{44}^c$ give a slightly better fit than $\braket{j_0}$, but since planar angular anisotropy is forbidden by symmetry until $l$ = 4, the resulting fit is close to isotropic for small $|\vec{Q}|$. Despite this, the direct space scalar magnetization density does not look isotropic. Instead an eight-fold symmetry is apparent within the honeycomb layers alternating between positive and negative values. Out of plane, two positive lobes lobed extend above and below the Ru center. This model also shows paramagnetic links between Ru chains along the $a$-axis with diamagnetic links between chains.

In the case of a trigonal compression along a $\{1$~$1$~$1\}$, the local symmetry changes to $\bar{3}m$ which permit four terms: $Z_{00}$, $Z_{20}$,  $Z_{40}$, and $Z_{43}^c$~\cite{coppens1997x}. While the best fit model for this symmetry is discussed in the main article, we note here that we also fit slightly reduced $\bar{3}$ symmetry, which in addition to the four terms for $\bar{3}m$ also allows a fifth term $Z_{43}^s$. In our case attempts to fit $C_3$ symmetry always fit $Z_{43}^s$ to be approximately zero within error and the model was reduced to that of $\bar{3}m$ up to $l$ = 4. This was robust against initial values of parameters and choice of local $z$-axis and suggests that deviations from a local $\bar{3m}$ symmetry are small for \rucl. 

The high temperature structure of \rucl\ is monoclinic, with the site symmetry of Ru ions being $C_{2}$. While the exact form of the low temperature is not as well understood, the strongly 1$^{st}$ order nature of the structural transition can lead to the coexistence of the low and high temperature structures. For the reason, we also investigated a monoclinic distortion symmetry. The relatively low symmetry $C_2$ allows nine terms in the multipole expansion up to $l$ = 4 ($Z_{00}$, $Z_{20}$, $Z_{22}^c$, $Z_{22}^s$, $Z_{40}$, $Z_{42}^c$, $Z_{42}^2$, $Z_{44}^c$, $Z_{44}^s$). Surprisingly the best-fit monoclinic structure bears strongly resemblance to the $D_{3d}$ best fit model in the honeycomb planes, with a distorted six-fold structure centered on Ru-ions observed in the plane. A slight compression of the lobes along the crystallographic $b$-axis along with a two-one alternation of the sign of the magnetization density distinguish the two. We also see that the monoclinic best fit multipole model has no change of sign in the magnetization density mirrored across the honeycomb plane.  The $C_2$ model shows an alternating positive and negative link within Ru chains along the $a$-axis, with little connection shown between chains.

While the best fit radial parameter ranges between multipole models, we see the out of plane extent is similar in the direct space scalar magnetization density. Aside from hints of direct overlap in these single ion calculations, we see that magnetization density extend relatively far toward Cl anions. This is also suggestive that hybridization effects are present in \rucl. 

Finally we note that while both Ru ions sit at identical 6$c$ Wyckoff positions, small deviations from a perfect octahedral coordination of the Cl separates the two Ru atoms into distinct but related sites. In our analysis, we also tried modeling the data using two separate form factors between the two Ru ions. Within our experimental data, we find no evidence of a noticeable difference between the two Ru sites, indicating that the deviations between the two Ru ions are negligible compared to our experimental uncertainty. Instead the lowest relative reduced $\chi^2$ is obtained from treating the two with a single magnetic form factor.

\end{document}